\documentclass[aps,prb,reprint,superscriptaddress,showpacs]{revtex4-1}
\usepackage{color}
\usepackage{graphicx}
\usepackage[american]{babel}
\usepackage[T1]{fontenc}
\usepackage{amsmath}
\usepackage{amssymb}
\usepackage{amstext}
\usepackage{amsthm}
\usepackage{latexsym}
\usepackage{verbatim}
\usepackage{natbib}
\usepackage{array}
\usepackage{color} 

\usepackage{ulem}   % For strikethrough

\def\214{Sr$_2$IrO$_4$}

\def\327{Sr$_3$Ir$_2$O$_7$}
\def\J{J$_{1/2}$}
\def\JJJ{J$_{3/2}$}

\def\TN{T$_{N}$}

\usepackage[draft]{todonotes}   % notes showed

\usepackage{xspace}

\begin{document}

\title {Evolution of the spectral lineshape at the magnetic transition in \214 and \327 }
%\vspace{1cm}

\author{Paul Foulquier}
\affiliation {Universit\'{e} Paris-Saclay, CNRS, Laboratoire de Physique des Solides, 91405, Orsay, France.}
\affiliation{Universit\'{e} Paris-Saclay, CEA, CNRS, SPEC, 91191, Gif-sur-Yvette, France}

\author{Marcello Civelli}
\author{Marcelo Rozenberg}
\affiliation {Universit\'{e} Paris-Saclay, CNRS, Laboratoire de Physique des Solides, 91405, Orsay, France.}

\author{Alberto Camjayi}
\author{Joel Bobadilla}
\affiliation {Departamento de F\'{i}sica, FCEN, UBA, and IFIBA, CONICET, Pabell\'{o}n 1, Ciudad Universitaria, 1428 Buenos Aires, Argentina}

\author{Doroth\'{e}e Colson}
\author{Anne Forget}
\affiliation{Universit\'{e} Paris-Saclay, CEA, CNRS, SPEC, 91191, Gif-sur-Yvette, France}

\author{Pierre Thu\'{e}ry}
\affiliation{Universit\'{e} Paris-Saclay, CEA, CNRS, NIMBE, 91191, Gif-sur-Yvette, France}

\author{Fran\c cois Bertran}
\author{Patrick Le F\`evre}
\affiliation {Synchrotron SOLEIL, L'Orme des Merisiers, Saint-Aubin-BP 48, 91192 Gif sur Yvette, France}

\author{V\'{e}ronique Brouet}
\affiliation {Universit\'{e} Paris-Saclay, CNRS, Laboratoire de Physique des Solides, 91405, Orsay, France.}

\begin{abstract}
  \214 and \327 form two families of spin-orbit Mott insulators with quite different charge gaps and an antiferromagnetic (AF) ground state. This offers a unique opportunity to study the impact of long-range magnetic order in Mott insulators. It appears to play a different role in the two families, as there is almost no change of the resistivity at the magnetic transition $T_N$ in \214 and a large one in \327. We use angle-resolved photoemission to study the evolution of the spectral lineshape through the magnetic transition. We use Ru and La substitutions to tune $T_N$ and discriminate changes due to temperature from those due to magnetic order. We evidence a shift and a transfer of spectral weight in the gap at $T_N$ in \327, which is absent in \214. We assign this behavior to a significantly larger coherent contribution to the spectral lineshape in \327, which evolves strongly at $T_N$. On the contrary, the \214 lineshape is dominated by the incoherent part, which is insensitive to $T_N$. We compare these findings to theoretical expections of the Slater vs Mott antiferromagnetism within Dynamical Mean Field Theory.
 \end{abstract}

\date{\today}

\maketitle

%=== Fig. 1 : Resistivity============================
\begin{figure}
%\centering
\includegraphics[width=1\linewidth]{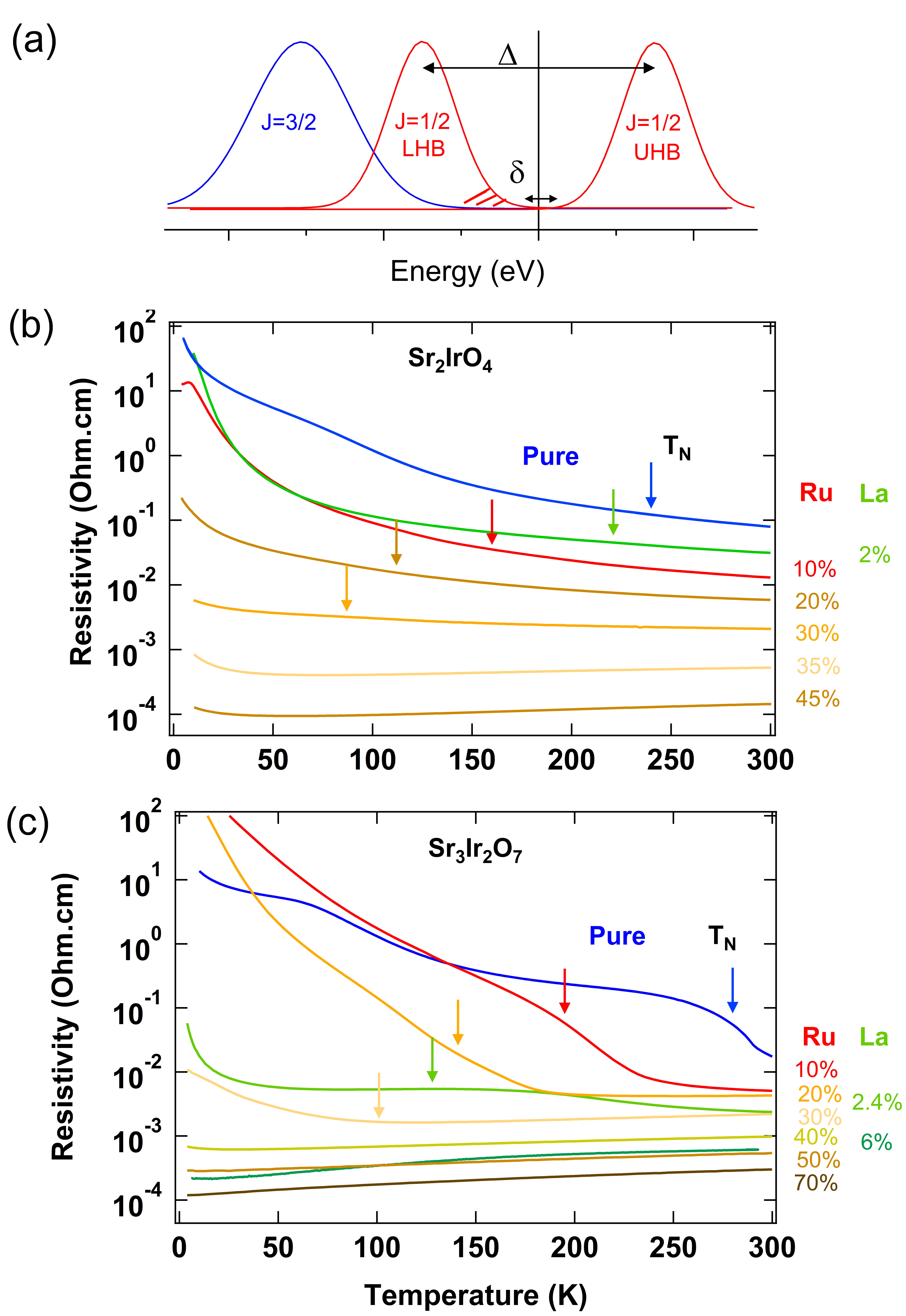}
\caption{(a) Sketch of the electronic structure expected for \214 and \327. The \JJJ~band is filled with 4 electrons, the \J~band is half-filled and split into Lower Hubbard Band (LHB) and Upper Hubbard Band (UHB) by $\Delta$. The gap between the tail of the bands is $\delta$. (b) Temperature dependence of the resistivity in \214 either pure (blue line), or doped with Ru (red-brown lines) or La (green line). The doping levels are indicated on the graph. \TN, determined by magnetic measurements, is indicated by an arrow. (c) Same as (b) for \327.} 
\label{Resistivity}
\end{figure}
%===============================

\section{Introduction}

The motion of one hole in an antiferromagnetic (AF) background is a central problem for many correlated systems, among which high temperature cuprate superconductors remain a hallmark. As moving one hole necessarily breaks some AF bonds, the charge and spin sectors become intimately intricated. The relative energy scale for charges are of the order of the charge gap $\Delta$, set by Coulomb repulsion U in a Mott insulator, independently of the magnetic order. AF order usually sets in at a lower temperature, depending on magnetic couplings J. Most treatments of the metal-insulator transition (MIT) are implicitely deep in the Mott insulating state and neglect the role of magnetic order. This is typically relevant for cuprates where U>>J. However, when the two energy scales become similar, closer to the MIT, the interplay between the two degrees of freedom may become quite complex \cite{HaoNatCom19,FratinoPRB17}. The role of AF long-range order on the MIT has been treated with DMFT theory in ref. \onlinecite{CamjayiPRB06} and predicts the coexistence of Mott-like and Slater-like excitations. To our knowledge, this has never been directly compared to experiments. 

Iridates offer a unique opportunity to carry over this comparison by tuning ($\Delta$, J) parameters. Indeed, the first two members of the Ruddlesden-Popper perovskite serie, \214 and \327, have quite significantly different charge gaps $\Delta$ at low temperatures, but similar magnetic transition temperature \TN. The two families can essentially be understood \cite{BJKimPRL08} as built from two filled \JJJ~bands and a half-filled \J~band, which is split by the electronic correlation opening a gap $\Delta$ (see sketch in Fig. \ref{Resistivity}a, J is the effective angular momentum). The Mott nature of the insulating state in \214 was supported by cluster-DMFT calculations\cite{MartinsPRM18}. \214 is built from single IrO$_2$ layers, stacked with SrO layers, and $\Delta\simeq0.6eV$ was estimated by optical spectroscopy \cite{MoonPRL08}, STM \cite{BattistiNatPhys17,ZhaoNatPhys19,SunPRR21} and ARPES using a small electron doping to visualize the whole gap \cite{BrouetPRB15}. For the bilayer version \327, the gap is smaller $\Delta\simeq0.2eV$ \cite{MoonPRL08,OkadaNatMat13,ZhaoScienceAdv21,BrouetPRB18_327}. The different $\Delta$ in the two families is essentially understood from the different effective dimensionality of the structure, leading to larger bandwidth for the bilayer compared to single layer \cite{MoonPRL08}. There are also some more subtle differences in the electronic structures. For example, interactions within the bilayer bands in \327 leads to a non-interacting semi-metallic electronic structure with an indirect gap of 0.05eV \cite{DeLaTorrePRL14}, which is simply enlarged by correlations \cite{BrouetPRB18_327,sup}. 

The two compounds display slight differences also in the AF magnetic structure. In \214 a transition takes place at \TN=240K to a canted in-plane AF state, where RIXS measured Heisenberg-like magnon dispersion over 0.2eV, characterized by J=0.06eV between first neighbors \cite{KimPRL12_214}. In \327, the magnetic transition at \TN=280K gives rise to an AF order with moments along the c-axis \cite{DhitalPRB13}. The magnetic interaction have similar order of magnitude (J=0.09eV), but the magnon dispersion is characterized by a large gap of 0.07eV due to pseudodipolar interactions in this geometry \cite{KimPRL12_327}. 

On one hand, the two compounds display insulating behavior in transport measurements above \TN, up to more than 600K\cite{KornetaPRB10,CaoPRB02}. This suggests that correlations, including short-range magnetic correlations, which persist above \TN \cite{ChenPRB15,ZhaoNatPhys19}, are responsible for the insulating properties, rather than the magnetic order. On the other hand, there is a rather strong temperature evolution in optical spectroscopy, characterized by weight appearing in the gap at high temperatures \cite{MoonPRB09,AhnSciRep16,XuBernhardPRL20}, indicative of bad metal properties. As the evolution is smooth and the transition temperature relatively high, it is difficult to disentangle the role of temperature and magnetic transition. These were tentatively attributed to the temperature dependence of polaronic excitations \cite{SohnPRB14}, but never fully clarified. More recently, Song et al. used Ru doping, which decreases \TN, to correlate a transfer of spectral weight with \TN~in \327 \cite{SongMoonPRB18}. Another optical study in Rh-doped \214 observed a transfer of spectral weight to a mid-infrared peak, which was interpreted as a spin-polaron feature from a one band Hubbard model \cite{XuBernhardPRL20}. From a different viewpoint, STM favors an inhomogeneous picture, where in-gap states are observed near dopants \cite{BattistiNatPhys17,WangNPJquantum19} or defects \cite{OkadaNatMat13} and may lead to a percolative-like MIT \cite{SunPRR21}. Angle-resolved photoemission could in principle go further by resolving the gapped structure in k-space. Its lineshape could help understand the nature of coherent and incoherent excitations and the possible emergence of in-gap states. However, there are few ARPES data available as a function of temperature and only for \327, either pure \cite{KingPRB13} or doped with La \cite{AffeldtPRB17_VsT} or 30\% Ru \cite{SongMoonPRB18}. 

\vspace{0.2cm}
\begin{figure}[tp]
%\centering
\includegraphics[width=1.0\linewidth]{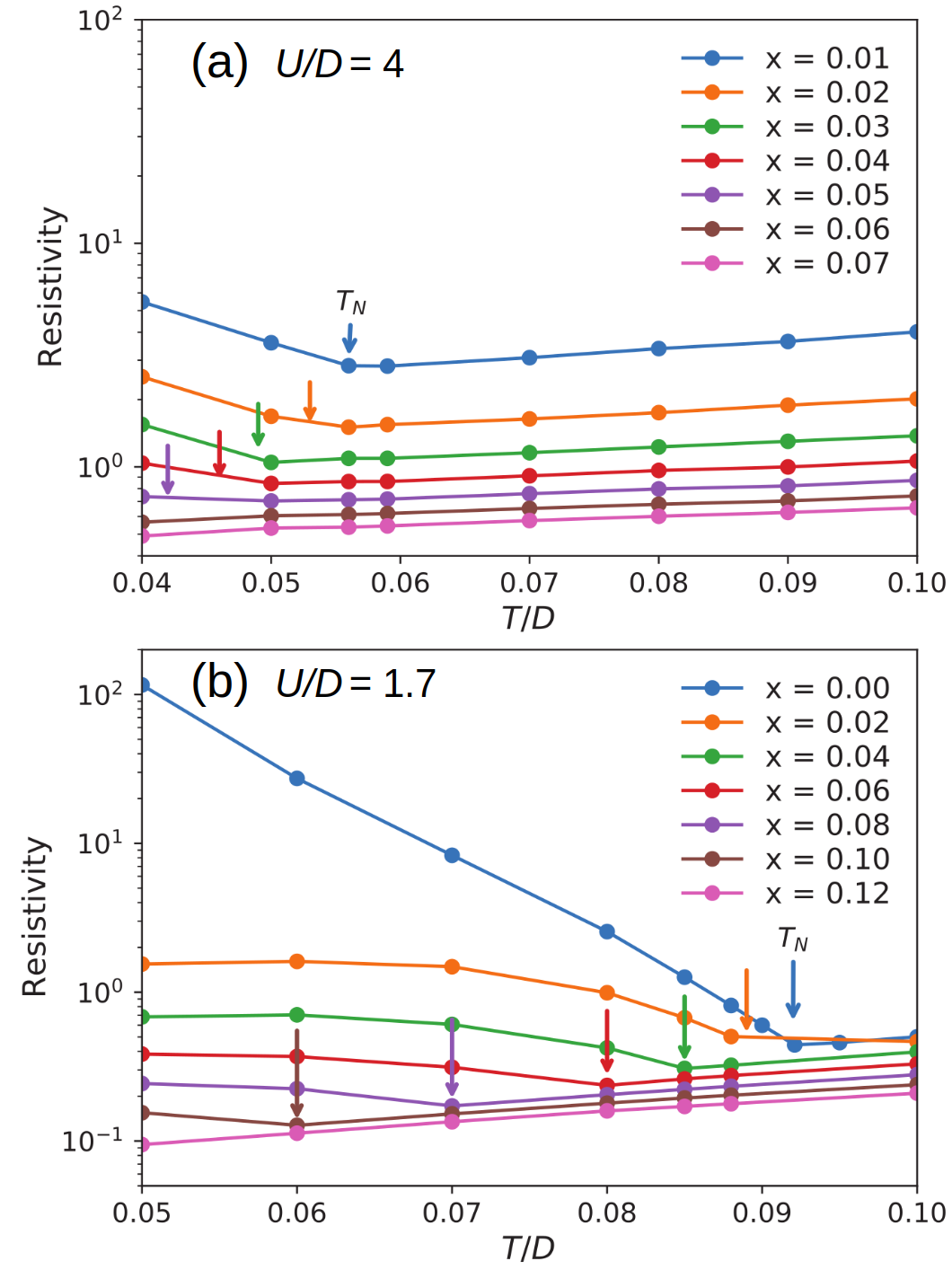}
\caption{(a) Temperature dependence of the resistivity obtained within DMFT
    for strong interaction $U/D=4$ for various doping x. The magnetic transition temperature $T_N$ is indicated
    by an arrow. (b) same as (a) for $U/D=1.7$. } 
\label{Resistivity-TH2}
\end{figure}

In this paper, we study systematically the evolution of the ARPES lineshape through \TN~in \214 and \327 and use La \cite{GeCaoPRB11,HoganPRL15} and Ru \cite{DhitalNatCom14,YuanPRB15,CalderPRB15} substitutions to tune \TN. La substitutes out-of-plane for Sr (we define x as Sr$_{1-x}$La$_{x}$) and induces electron doping. This leads to the reduction of the magnetic order, which vanishes around x=0.04 \cite{GeCaoPRB11,HoganPRL15}. On the contrary, Ru substitutes in-plane for Ir (Ir$_{1-x}$Ru$_{x}$) and, although Ru has one less electron than Ir, it seems there is an electronic phase separation at early dopings, so that Ru dilutes the magnetic state rather than dopes it \cite{DhitalNatCom14,YuanPRB15,CalderPRB15,BrouetPRB21}. A high concentration around x=0.35 is required to suppress the magnetic order and induce a metallic state. 

We shall qualitatively compare our results with reference theoretical results obtained by Dynamical Mean Field Theory on the doped
  Hubbard Model. A detailed quantitative comparison with iridates would require employing more realistic material approaches, which can include short ranged spacial correlations (e.g. the cluster extension of DMFT), spin-orbit coupling, multi-bands effects and possibly weakly correlated bands within the ab-initio framework.  However such studies, which demand future developments \cite{MartinsPRM18,MoutenetPRB18,ZhangPRL13,JeongPRL20}, are beyond the scope of this paper which focus on the general properties of the antiferromagnetic Slater to Mott crossover. We shall consider two different interaction regimes: the rather weakly correlated one ($U/D$= 1.7), which is dominated by the Slater AF mechanism, and the strongly correlated one ($U/D$= 4), dominated by the Mott localization mechanism. We shall show that as a matter of facts many physical properties of \327 can be described with the former Slater regime, while the
  \214 well fits the latter Mott regime. This is especially true for transport, but also for the ARPES spectral lines,
  provided we make some assumption on features related to the chemical substitution and disorder, which can broaden the spectra and affect the Fermi level position.
Our experimental-theoretical comparison shows that Ir-based oxides can provide a unique platform to study non-trivial correlation phenomena, like the
  evolution from weak to strong correlation, the interplay of Slater magnetism and Mott localization and the effects of
  doping, temperature and disorder.

\section{Methods}
Single crystals of \214 and \327 were grown by a standard flux growth technique. High-purity powders of SrCO$_3$ (99.995\%), IrO$_2$ (99\%), RuO$_2$ (99.9\%) were dried, weighed, and mixed in a glove box under argon with SrCl$_2$ (99.5\%) flux. The mixture was loaded into a platinum crucible covered with a platinum tip, under ambient atmosphere, and heated in a muffle furnace. For \214, we used ratios 2:1:10, heated up to 1300 $^{\circ}$ C and then slowly cooled down at a rate of 10$^{\circ}$ C/h to 800$^{\circ}$ C. For \327, we used ratios 3:2:5, heated at a rate of 190$^{\circ}$ C/h up to 1100$^{\circ}$ C for 6 hours and then slowly cooled down at a rate of 10$^{\circ}$ C/h to 600$^{\circ}$ C, at which temperature the furnace was turned off. Deionized water was used to dissolve the SrCl$_2$ flux and extract the single crystals. The crystals were platelets with larger dimensions between 0.3 and 2 mm, and with the smallest dimension along the [001] direction. The exact composition of each studied sample has been determined via energy-dispersive X-ray spectroscopy (EDS) measurements in several spots of the surface of several crystals from the same batch. The structure was further refined by x-ray diffraction. The results for nine samples of Sr$_{3}$(Ir$_{1-x}$Ru$_{x}$)$_2$O$_{7}$ with x in the 0-0.78 range are given as supplementary material.

ARPES experiments were carried out at the CASSIOPEE beamline of SOLEIL synchrotron, with a SCIENTA R-4000 analyser, 100 eV photon energy and an overall resolution better than 15meV.

DMFT calculations where performed on the Hubbard Model, the reference playground to study correlated phenomena
\cite{GeorgesRMP96}.
  The model has the typical semi-circular density of states of bandwith $D$, which fixes the energy unit
  throughout the paper. The DMFT is implemented by means of the continuous time diagrammatic Quantum Monte Carlo
  \cite{GullRMP11}.
  Spectra are obtained via analytic continuation of the one particle propagator performed by the Maximum entropy method \cite{Levy17}.
  DMFT allows to unbiasedly access the paramagnetic insulating and metallic states, as well as the ordered antiferromagnetic
  insulator\cite{CamjayiPRB06,FratinoPRB17}. Here we shall study how these states evolve and compete upon doping the system with holes.  
	
  %=== Fig. 3 : Overview of Lineshape ============================
\begin{figure*}
%\centering
\includegraphics[width=0.8\linewidth]{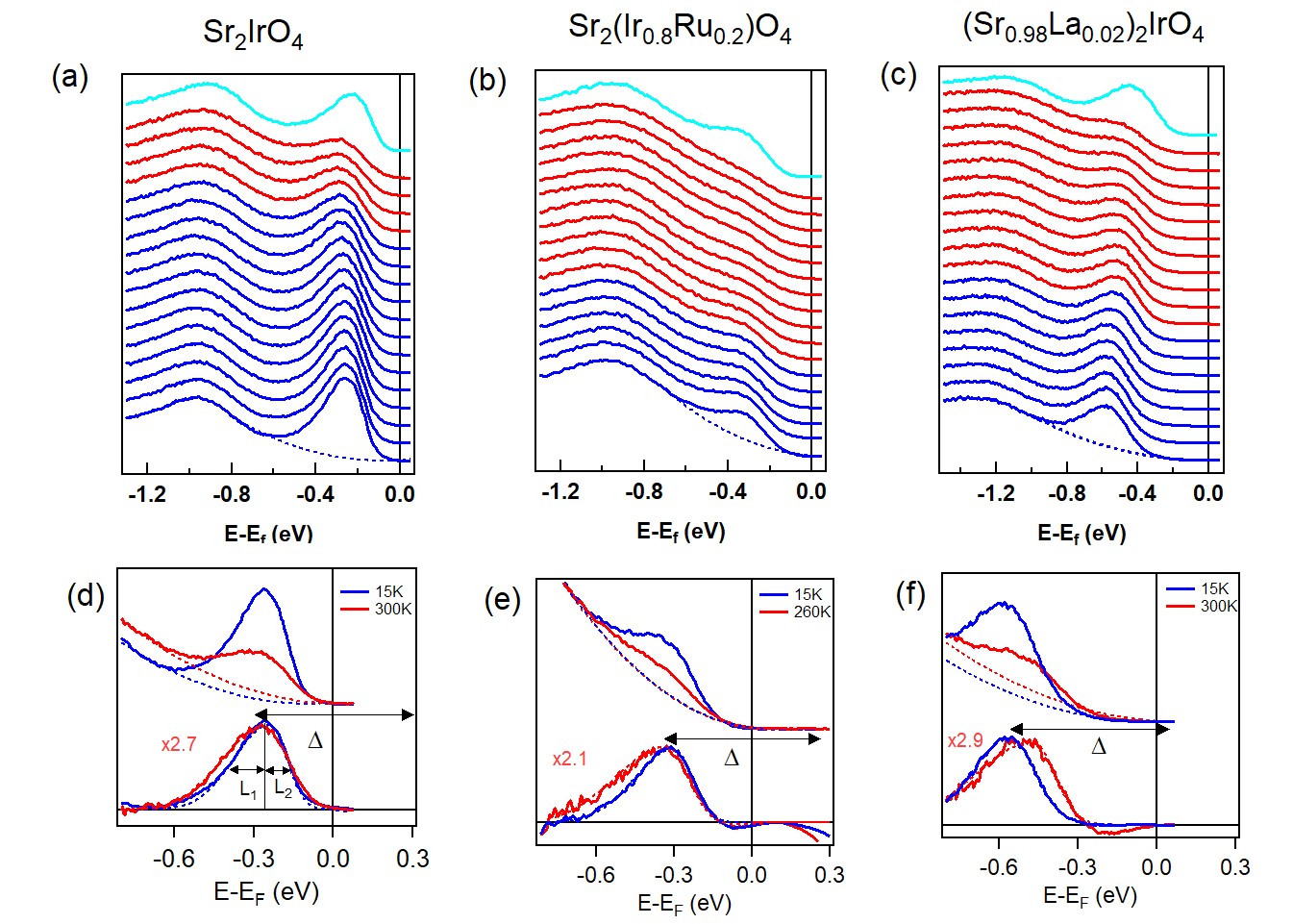}
\caption{(a-c) Stacks of EDC spectra at X in the three indicated compounds as a function of temperature (from lowest in blue, 20K, to highest in red, 300K). The top spectrum (light blue) is taken at 20K after a temperature cycle. The blue lines are in the magnetic state, the red ones in the paramagnetic state (\TN=240K for the pure, 150K doped with 20\%Ru and 200K for 2\% La). Dotted lines indicate the baseline used in the fit. (d-f) Top : zoom on the spectra superimposed for low and high temperatures, with the baselines as dotted lines. Bottom : corresponding spectra with baseline subtracted. The arrow indicates the size of the gap $\Delta$ expected at X (0.6eV in each case). The two spectra are normalized to their maximum for easier comparaison of the lineshape (high temperature spectra are magnified by the indicated amounts). The fit used to extract position and width in Fig. \ref{VsT} are shown as dotted lines (often indistinguishable from the raw data). The model chosen for the fit (an asymmetric Gaussian with width $L_1$ and $L_2$) is also presented. } 
\label{Resub_Fig3_214}
\end{figure*}
%===============================

\vspace{0.5cm}

\section{Resistivity behavior}

A clear indication of the different role of the magnetic transition in the two families is already evident in the evolution of resistivities plotted in Fig. \ref{Resistivity}. The arrows indicate \TN, as determined from magnetic measurements (by SQUID in \214 and neutrons in \327 \cite{DhitalNatCom14,HoganPRB16,sup}). There is almost no change in \214 at \TN, while there is a clear anomaly in \327 leading to a more conducting state in the paramagnetic regime. 

The resistivity does not follow a simple activated behavior on the full temperature range. Fitting to an activated law between 100 and 200K gives a gap of the same order of magnitude in the two systems, $\delta\simeq$60meV. This is much smaller than the gap $\Delta$ previously evaluated. It can be understood as the smallest energy distance between the tail of the peaks (see Fig. \ref{Resistivity}a). This emphasizes the importance of in-gapped low energy states in these systems, but also that limited information on the evolution of $\Delta$ can be extracted from resistivities alone. 

This qualitative behavior is in good agreement with what is expected within a Slater-to-Mott antiferromagnetic crossover.
  To show this point, we plot in Fig. \ref{Resistivity-TH2} the resistivity vs temperature obtained
  by DMFT for two values of the interactions strength,
  $U/D= 4.0$ (panel a), which is deep in the strongly correlated Mott regime,
  and $U/D= 1.7$ (panel b), which is in the weakly correlated regime. The system is lightly doped, up to $x=0-10\%$.

  These theoretical curves display a qualitative behavior in line with what is observed in the experimental curves described above
  in Fig. \ref{Resistivity}. Namely, in the correlated regime, alike to \214,
  the resistivity vs $T$ curve is rather flat and 
  does not display any anomaly in correspondence to the antiferromagnetic transition temperature $T_N$. In sharp
  contrast, the weakly correlated regime looks like \327, clearly displaying an anomaly in the curve in line with $T_N$.
  These features disappear at higher doping level in the absence of magnetic transition.
		
		In tracing the comparison between these experimental and theoretical resistivity curves, some caveats must be taken
    into consideration. In the strongly correlated regime $U/D=4$, the system immediately becomes metallic upon doping,
    though the metallic character is very weak (for example, the absolute value of the resistivity on a doped state curve for $U/D=4$ is an order of magnitude higher than the resistivity of the
    weakly correlated $U/D=1.7$ case).  In a real material, such a state would likely display insulating-like properties. Disorder, not taken into account in our theory, can play an important role and localize a small number of carriers. This was observed for instance in Rh-doped \214 \cite{LouatPRB18}. Overall, our theory-experiment comparison enforces 
    the interpretation that \214 is deep in the Mott state, while in \327, both correlation and Slater antiferromagnetism play a key role.

\section{Overview of ARPES lineshapes}
\begin{figure*}
%\centering
\includegraphics[width=0.8\linewidth]{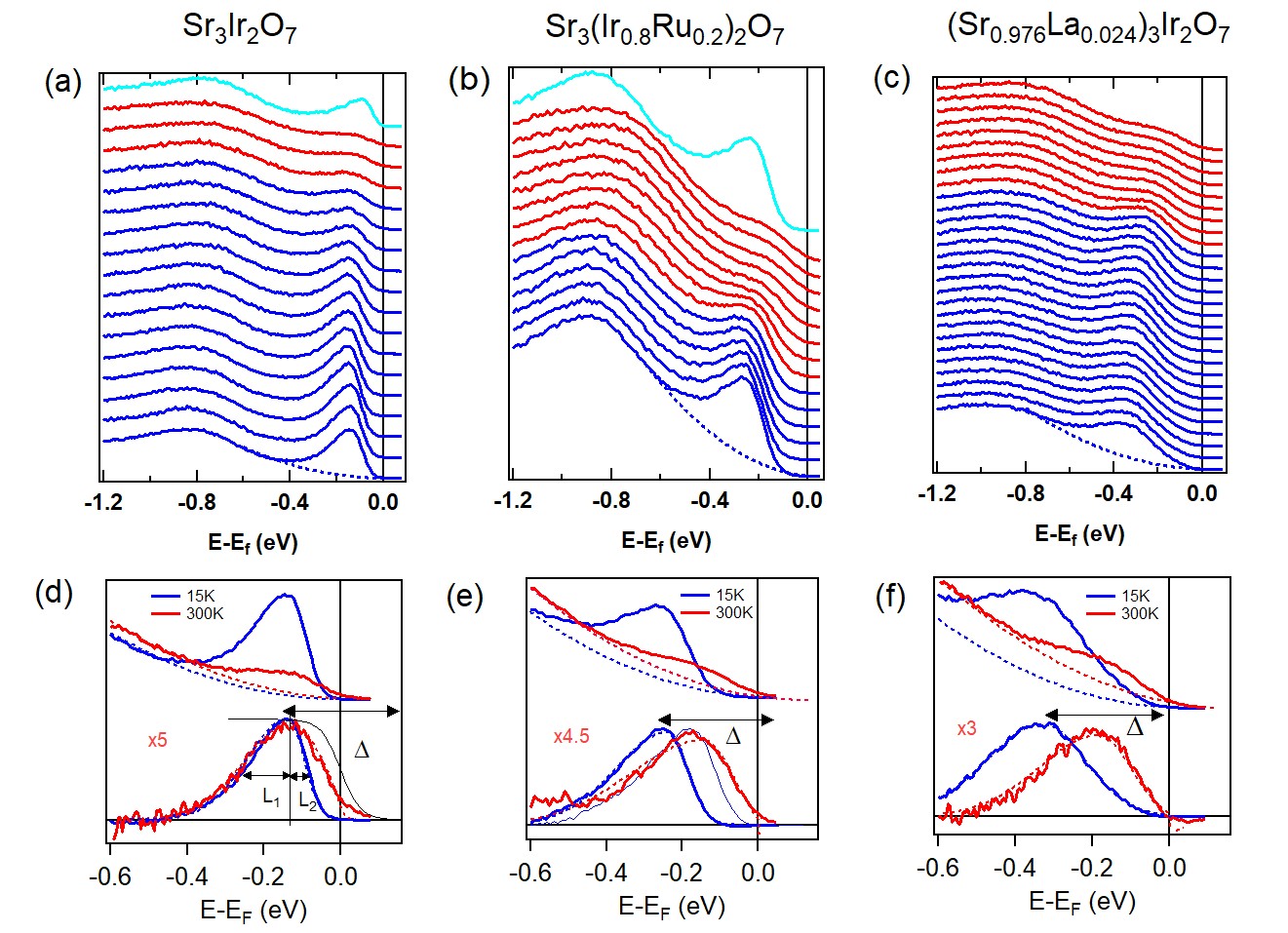}
\caption{Same as Figure \ref{Resub_Fig3_214} for \327 family. \TN=280K for the pure, 180K doped with 20\%Ru and 130K for 2.4\% La. The arrow indicating the size of the gap $\Delta$ measures 0.3eV. In (d) a Fermi-Dirac step at 300K is shown as thin black line. In (e), the low temperature spectrum shifted to the high temperature one position is shown as thin blue line.} 
\label{Resub_Fig3_327}
\end{figure*}

Fig. \ref{Resub_Fig3_214} and \ref{Resub_Fig3_327} present ARPES Energy Distribution Curves (EDC) as a function of temperature taken at the top of the \J~band, located at the X point of the reciprocal space (see supplementary material \cite{sup} for a sketch of the electronic structure). The position of the magnetic transition is indicated by the change of line color (blue to red). In each family, we examine three cases : pristine compound (\TN=240K for \214 and 280K for \327), doped with 20\% Ru (\TN=150K and 180K respectively) and with a few percent La [\TN=200K (2\% \214) and 130K (2.4\% \327)]. More dopings are presented in supplementary information, these ones illustrate the universality of the behavior. Two bands can be observed in each EDC, \J~around -0.2/-0.3 eV binding energies and \JJJ~around -0.8/-1eV \cite{BrouetPRB15,BrouetPRB18_327}. At the bottom of Fig. \ref{Resub_Fig3_214} and \ref{Resub_Fig3_327}, we show superimposed spectra at low and high temperatures. To isolate the shape of the \J~peak, we subtract a parabolic baseline (bottom, see supplementary material \cite{sup} for more details). The spectra are scaled to their total area. The high temperature spectra is magnified by the indicated number to better compare the lineshapes.

%=== Fig. 3 : Overview of Lineshape ============================
%===============================

%Fig. \ref{Lineshape_214}(a-b) gives an overview of the changes of the spectra in undoped \214 and \327 across the magnetic transition. 
For the pure compounds, a  gap $\Delta$ opens up in \J~at X. The Fermi level is roughly in the middle of the gap $\Delta$ estimated by STM and optics for the undoped systems. The peaks are quite broad (0.2eV at half maximum for \214 (Fig. \ref{Resub_Fig3_214}d) and 0.15eV for \327 (Fig. \ref{Resub_Fig3_327}d) and the distance between the tail of the peaks will be significantly smaller than the peak to peak positions, in agreement with the smaller gap $\delta$ dominating the resistivity, as indicate in Fig. \ref{Resistivity}a. For a quasiparticle (QP) excitation, the ARPES lineshape should be lorentzian-like with a width given by the QP inverse lifetime. However, the peak here is rather gaussian-like, asymmetric and much broader than what would be expected for a QP. This situation is common to many insulating oxides, as cuprates \cite{KimPRB02} or manganites \cite{DessauPRL98}. This suggests a composite nature of the line, where the spectrum is the envelope of a distribution of excitations. %Asymmetry of ARPES lineshape in insulators can often be attributed to plasmonic-like losses or more generally to any incoherent tail. 
A possible origin is the formation of polarons \cite{PerfettiPRL01,DessauPRL98,KMShenPRL04}, as was actually proposed to explain the ARPES linewidth in \327 \cite{KingPRB13}. In this case, the shape of the peak is fixed by the strength of electron-phonon coupling, asymmetric for low couplings and gaussian for higher ones. As there is no obvious reason why the electron-phonon coupling should be different in \214 and \327, this picture does not explain easily the much more asymmetric lineshape of \327. Also, we will see that the lineshape changes at \TN, whereas no strong evolution of the electron-phonon coupling would a priori be expected there. Indeed, Raman experiments, which are sensitive to the phonon renormalization due to pseudospin-lattice coupling, do not evidence large changes at \TN \cite{GretarssonPRB17}.

%=== Fig. 4 : Lineshape analysis ============================
\begin{figure*}
%\centering
\includegraphics[width=1\linewidth]{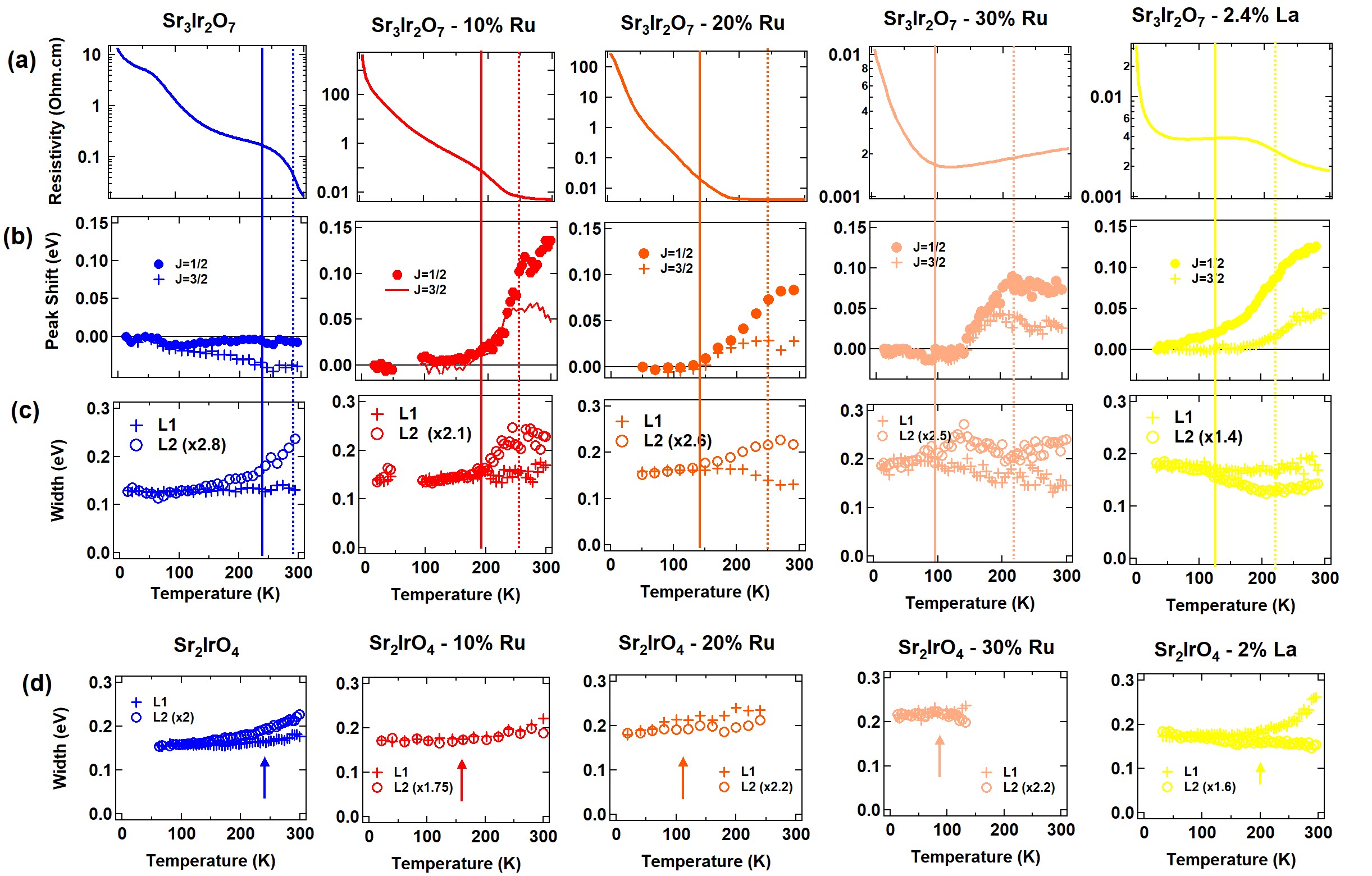}
\caption{(a) Resistivity in (doped) \327. The vertical lines indicate the width of the magnetic transition, as seen by neutron experiments (see text). (b) Shift of the ARPES peak position in (doped) \327 with respect to its low temperature value, for \J~at X (closed symbol) and \JJJ~at $\Gamma$ (crosses). Raw data are shown in the supplementary material. (c) The two widths $L_1$ (towards high binding energy) and $L_2$ (towards low binding energy) in (doped) \327 extracted from a fit of the \J~peak at X to an asymmetric gaussian. $L_2$ is magnified to match $L_1$ value at low temperatures, as indicated. (d) Same as (c) for (doped) \214. $T_N$ (arrow) is defined here by the onset of the ferromagnetic signal in SQUID measurements \cite{sup}.}    
\label{VsT}
\end{figure*}
%===============================

When the samples are lightly doped, the gap essentially does not change (as will be justified later), but the Fermi level moves inside the gap, as a result of filling of the first available states. In \214 (Fig. \ref{Resub_Fig3_214}), there is no big change of the lineshape between high and low temperatures, except for a slight broadening on both sides. In particular, there is no sudden shift at \TN, implying that the gap does not suddenly close. The \J~peak intensity however is strongly supressed at high temperature. This effect is only partially reversible and it is difficult to disentangle the role of the temperature increase and of \TN~in this intensity loss. 

On the contrary, in pure \327, there is a clear deformation of the spectrum, which extends towards $E_F$ (see Fig. \ref{Resub_Fig3_327}). The leading edge shifts up by 50meV, but remains away from $E_F$. A comparison with a Fermi-Dirac distribution at 300K (black line) suggests a remaining "pseudogap" of 50meV in Fig.\ref{Resub_Fig3_327}d. Nevertheless, some residual density appears at $E_F$, which is consistent with a bad metallic character. The peak maximum itself has not moved significantly, ruling out a sudden collapse of the gap at \TN. This comparison implies that the difference between the two systems is more complex than a gap simply closing in one case and not the other. It seems rather related to a transfer of weight in the gap for \327 that is not present for \214. In doped \327, there is in addition a clear shift of the spectra towards $E_F$, both for \J~and~\JJJ, bringing weight at the edge of $E_F$. As~\JJJ~should not be affected by a gap closure in~\J, this behavior suggests a shift of $E_F$ inside the gap. To compare the lineshape, we shift the low temperature spectra to the high temperature one for 20\% Ru (thin blue line, shifted up by 70meV). This reveals a similar change of the lineshape as for the pure, with a characteristic extension of the spectrum towards $E_F$. The comparaison is more difficult for the La case, where the low temperature spectrum is broader than the other ones, especially towards $E_F$. This is probably due to some distribution in La content, which hides the intrinsic lineshape at low temperatures.

\section{Change at $T_N$}

%=== Fig. 5 : Shift vs T ============================
\begin{figure*}
%\centering
\includegraphics[width=0.95\linewidth]{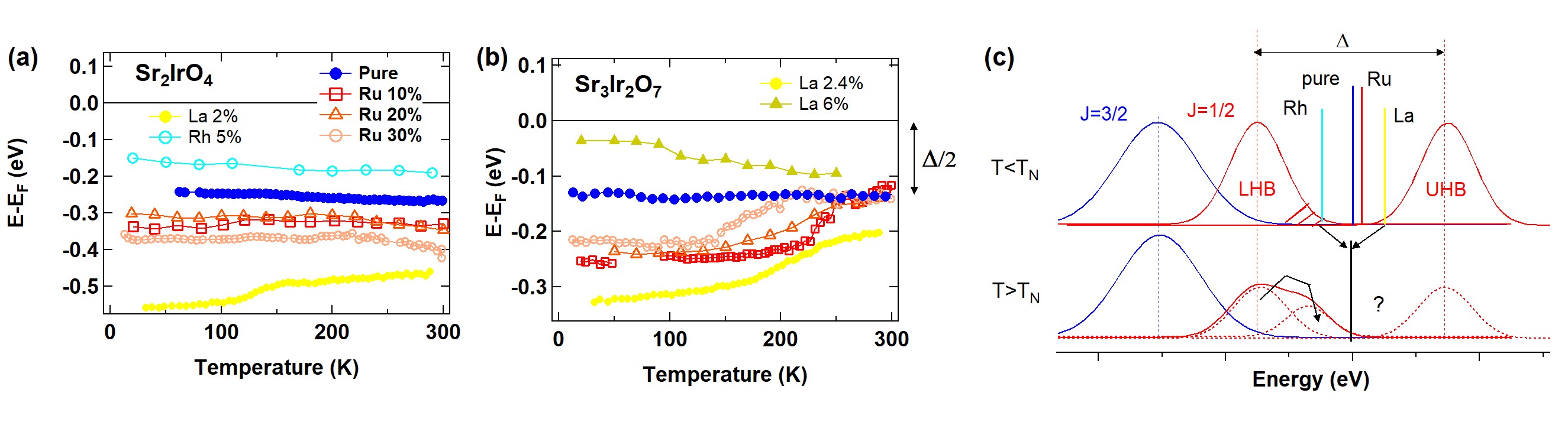}
\caption{Peak position as a function of temperature of the asymmetric gaussian used to fit the \J~EDC at X in \214 (a) and \327 (b) for the indicated dopings. (c) Sketch of the evolution at $T_N$, involving a transfer of spectral weight from the Hubbard bands to in-gap states. The initial position of $E_F$ for the different dopings is sketched by color lines on the top graph, which all move to the middle of the $\Delta$ gap above $T_N$. The detail of the bands above $T_N$ cannot be known from ARPES. }    
\label{shift}
\end{figure*}
%===============================

We now study how these temperature evolutions correlate with the magnetic transition. The magnetic transition is quite broad in doped iridates, probably proceeding through some phase separated region \cite{DhitalNatCom14}. For \214, we define $T_N$ by the onset of the ferromagnetic signal in SQUID measurements \cite{sup}. This is not possible for \327, where this signal is very weak, and we rely on neutrons scattering performed in Ru \cite{DhitalNatCom14} and La \cite{HoganPRL15} doped \327. We define two temperatures limiting the magnetic transition, $T_{N,1}$ where magnetic Bragg peaks appear, and $T_{N,2}$ where their intensity saturates. Experimentally, it is this latter value that corresponds to the anomaly of resistivity, reported as \TN~on Figure \ref{Resistivity}. To characterize the temperature dependence of the lineshape, we fit spectra at each temperature to an asymmetric gaussian, as shown in Fig.~\ref{Resub_Fig3_214}d and \ref{Resub_Fig3_327}d. This emphasizes the key evolution we have described and limits the number of parameters (more examples of fits are shown in the supplementary material \cite{sup}). 

We first focus on \327 where there are clear changes. In Fig. \ref{VsT}a, we observe that the change in resistivity starts at $T_{N,2}$ (solid line) and evoluates until roughly $T_{N,2}$ (dotted line). Similarly, the shift of the peak position of \J~at X, shown in Fig. \ref{VsT}b, is precisely tied to the magnetic transition, it starts at $T_{N,2}$ and saturates above $T_{N,1}$. We compare it to the shift of \JJJ~observed at $\Gamma$ (details are given in supplementary \cite{sup}) and find that it exhibits a very similar temperature evolution, although smaller by a factor 2. This suggests that the shift is at least partly due to a motion of $E_F$ in the gap and not simply to a change in the energy scales $\Delta$ and $\delta$. 

Interestingly, the lineshape also changes at the magnetic transition. In Fig. \ref{VsT}(c), we present the two widths of the asymmetric gaussian, $L_1$ (towards high binding energies) and $L_2$ (towards $E_F$), scaled to the low temperature value. We find that they are roughly constant below $T_{N,2}$, but $L_2$ increases above $T_{N,2}$, strongly diverging from $L_1$, which remains constant or even decreases. This is in agreement with the evolution described in Fig. \ref{Resub_Fig3_327}, where only the low energy side of the peak changes at high temperature. This however further demonstrates that this evolution is triggered by magnetic ordering. A first possibility would be that this reflects a distribution of positions arising above $T_N$. As we have seen that the transition is quite inhomogeneous, it has to be considered. However, a narrower lineshape would then be expected again, when the sample has fully transited, contrary to our observation (the linewidth saturates but remains broad above $T_{N,1}$). Therefore, although disorder certainly plays a role, it cannot explain the broadening above $T_N$. We then assume that spectral weight is transferred in the gap, deforming the lineshape towards $E_F$. 

This behavior (both the shift and the spectral evolution) is intrinsic to \327. In Fig. \ref{VsT}(d), we give a similar view of the evolution of the widths in \214 (we could not fit reliably the widths at high temperature for high Ru dopings, the spectral weight becoming too small to be separated from the background). They do not point to a systematic change of lineshape at $T_N$. The width broadens moderately with increasing temperature, but it is sometimes $L_1$ that is largest (La doped) or $L_2$ (pure) or they remain similar (Ru doped), with no clear anomaly through $T_N$.  

\vspace{0.2cm}
In Fig. \ref{shift}(a-b), we summarize the evolution of the \J~position in the two families. For \214, there is no particular shift at \TN. In the case of La, we observe a small shift with temperature, but it is not happening at $T_N$ (200K in this case) and it could be due to the formation of defect states in the gap moving $E_F$ to a new position. Indeed, it is not completely reversible (see Fig. \ref{Resub_Fig3_214}c). For completeness, we add the case of a small hole doping of \214, obtained by 5\% Rh \cite{LouatPRB18}  (\TN=170K), which confirms the absence of shift at $T_N$ in \214 family.

For \327, the shift is significant for all doped compounds and Fig. \ref{shift}(b) further reveals that all peaks seem to converge to the position of the pure at high temperature, independently of the doping. For the pure compound at low temperature, this position is well understood as corresponding to a Fermi level in the middle of the Mott gap, at $\Delta$/2. This convergence then means that the Fermi level is fixed at $\Delta$/2 at high temperatures, for all dopings. This also implies that there remains a Mott-like energy scale $\Delta$ above \TN~for both families. 

Fig. \ref{shift}(c) summarizes our understanding of the evolution. At low temperatures, the position of the Fermi level is fixed by the doping. It is either near the middle of the gap (pure compound), at the tail of UHB for electron doping (La case) or LHB for hole doping (Rh case). For more disordered situations (Ru case), it is found at intermediate position. Above $T_N$, the Mott-like gap $\Delta$ is not closing suddenly, neither for \214, nor for \327. There is a remaining LHB at $\Delta$/2 in both cases, which still dominates the spectral weight. However, spectral weight is transferred in the $\Delta$ gap, in a much more pronounced way for \327 than \214, and this transfer starts and stops over the width of the magnetic transition. This transfer fixes the position of the Fermi level at the center of the $\Delta$ energy scale, overruling the previous disorder/doping dependent position and producing a shift of the spectra in doped cases. By comparison, the absence of shift in \214 appears as a sensitive sign that there is no significant change of the in-gap structure.

To distinguish the "filling" of the Mott gap, from a "closure", we add the case of a higher La doping of 6\% in \327, where the metallic state is realized \cite{DeLaTorrePRL14,BrouetPRB18_327} (dark yellow triangles in Fig. \ref{shift}(b)). Clearly, the peak is much closer from $E_F$ at low temperatures in this metallic compound. We will further show in Fig. \ref{Fit} that the lineshape is also completely different, with a narrow peak.

\section{Discussion}
%===============================
\begin{figure}
%\centering
\includegraphics[width=1.0\linewidth]{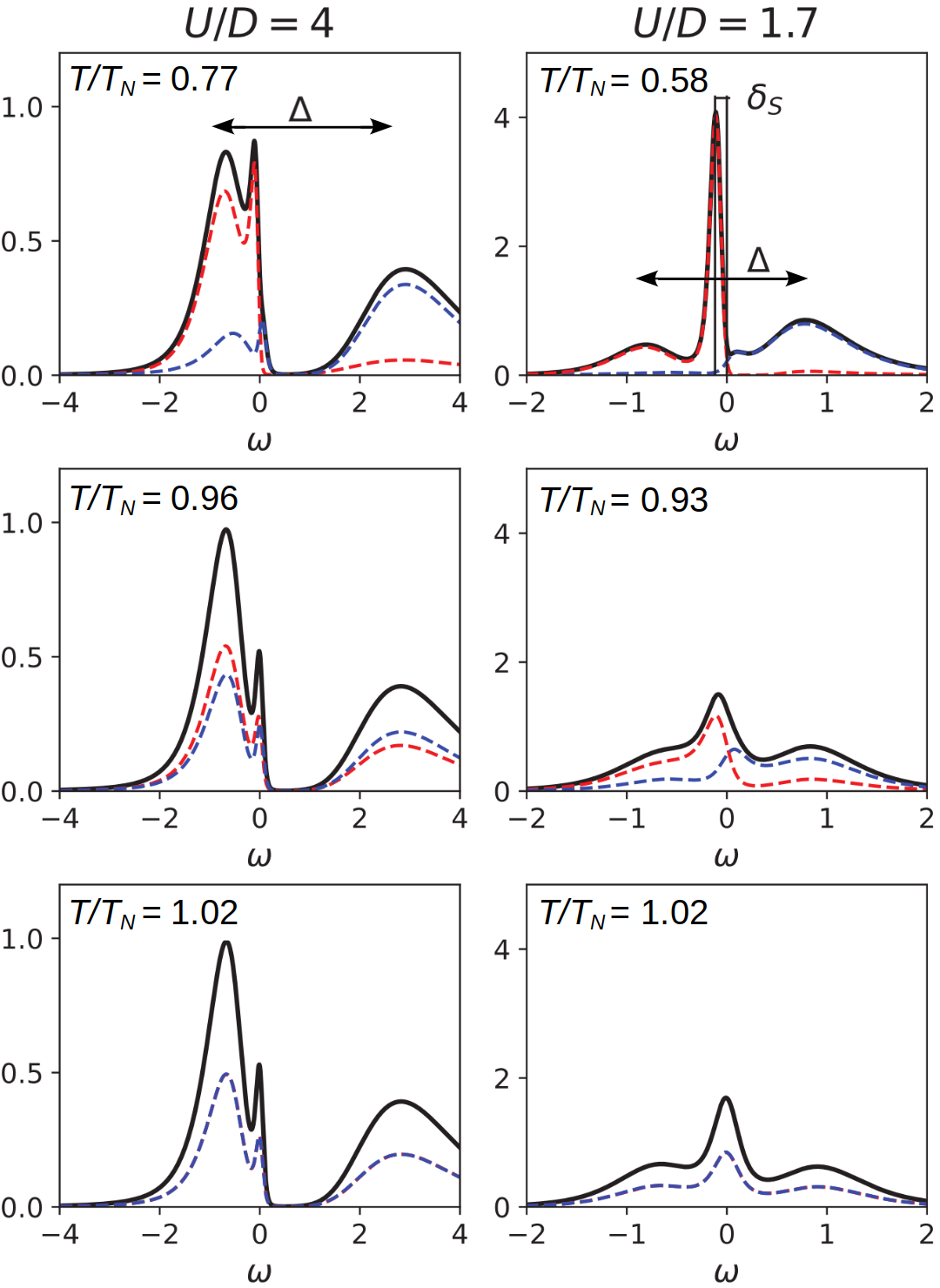}
\caption{Temperature evolution of antiferromagnetic DMFT spectra (from top to bottom) for the strongly correlated Mott case ($U/D=4$, left hand side) and the
    weakly correlated Slater case ($U/D=1.7$, right hand side). In the strongly correlated case $U/D=4$, a Mott energy scale $\Delta$ (separating the Hubbard bands)
    remains well defined and line-shapes do not essentially move. Antiferromagnetism is restored
    by the recovering of equal spin up (red) and spin down (blue) spectral weights. For weak correlation $U/D=1.7$, though Hubbard bands remain visible across T$_N$,
    an antiferromagnetic gap $\delta_S$ closes at the Fermi level across T$_N$, inducing a shift of spectral weight to the Fermi energy,
    as expected within a Slater mechanism. This forces the Fermi level to be located at the middle of $\Delta$. 
} 
\label{Fig6_DMFT}
\end{figure}
%========================================================

%=== Fig. 7 : Fit of lineshapes ============================
\begin{figure}
%\centering
\includegraphics[width=1\linewidth]{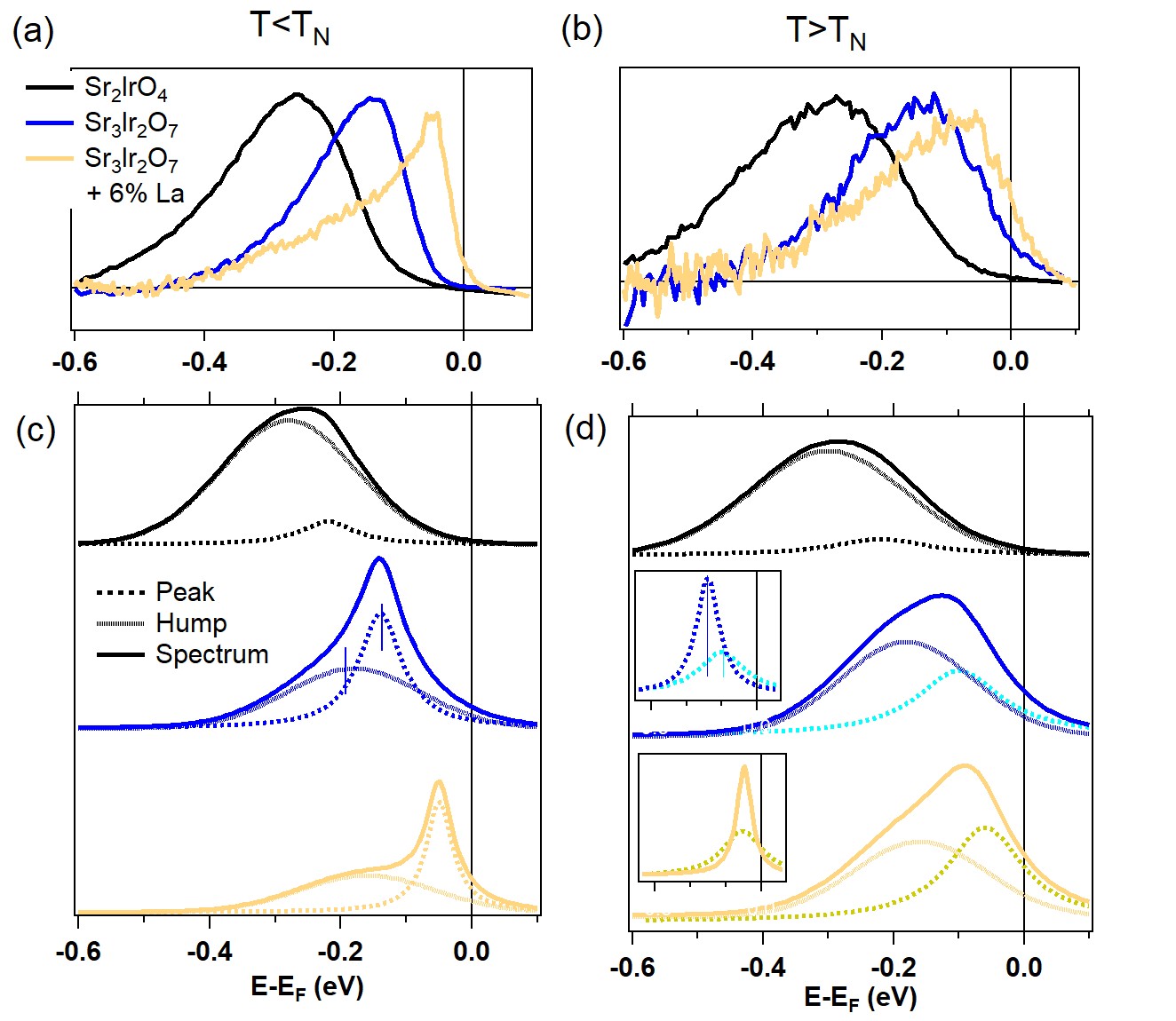}
\caption{(a) ARPES lineshape at 15K for \J~(EDC at X with subtracted background) for \214 (black), \327 (blue) and metallic \327 (doped with 6\% La, orange). (b) Same at 300K. (c) Simulation of the low temperature spectra with a peak and hump structure (see text). (d) Same for high temperature spectra. Insets show the evolution in temperature of the peak contributions.}    
\label{Fit}
\end{figure}

We now compare these experimental spectra with the theoretical expectations
  in a Mott vs Slater scenario. For this purpose we show in Fig. \ref{Fig6_DMFT}
    the DMFT spectra of the Hubbard Model at small doping.
  On the left column we display the strongly correlated Mott regime ($U=4D$), on the right
  one the weakly interacting Slater regime ($U=1.7D$), for increasing temperature (from top to bottom).
  The system is at small hole doping, $x=2.5 \%$ and $x=3.5\%$,
  for strong and weak correlations respectively. Spectra put into evidence the sharp difference between the Mott
  and the Slater mechanism closing of the gap above $T_N$.

  In the Mott regime a Mott energy scale is well defined at low
  temperature, marked as $\Delta$ on the top left panel, separating the LHB from the UHB.
  Upon doping, a small QP peak appears at the Fermi level, at the upper edge of the LHB. The AF is evident from the strong spectral weight differentiation
  between the spin up (red dashed line) and spin down (blue dashed line) species.
  This weak QP peak feature in the strong correlation regime is likely to be washed away by disorder
  and impurity effects in a real material. Indeed, we have seen with Figure 1 that lightly doped \214 is not immediately metallic, contrary to this prediction, and that $E_F$ is not fixed to the upper of LHB, but remains within the gap, at a position depending on doping and/or impurities. The key point is that upon increasing temperature, the spectra
	are little affected, and AF is suppressed only by recovering the balance between the spin up and spin down spectral weights.

  In the Slater regime, we can still identify a Mott energy   scale $\Delta$, separating the LHB and the UHB, though, with respect to the correlated Mott regime, this scale is now renormalized to a smaller value than the onsite interaction $U$. Much more pronounced QP peaks appear now at the edges of the Hubbard bands and they define a small gap $\delta_S$ around the Fermi level. This gap is this time opened by the AF mechanism (besides the spin up and spin down spectral weight differentiation),
  separating spin up and spin down QP-like peaks. This is
  the key difference with respect to the Mott regime, for which the QP peak, though weak, is at the Fermi level
  even at small doping and T<\TN. Upon increasing the temperature, while the Mott $\Delta$ scale is only slightly affected, the low $\delta_S$ energy scale collapses at
  the transition temperature $T_N$. At the same time the spin up and spin down spectral
  weight imbalance disappears. The distinctive feature is that with 
  the closing of the $\delta_S$ energy scale, a sharp increase of spectral weight appears at the Fermi level explaining the "anomaly" observed in the resistivity.  

We now attempt a direct comparison between the experimental spectra and the two theoretical scenarios. In Fig. \ref{Fit}(a-b), we compare three different ARPES lineshapes (a baseline has been subtracted) for large gap $\Delta$ (\214), smaller gap (\327) and a metallic case (\327 with 6\%~La). We find that the lineshape at low temperature in \214 ($L_1$=0.15 and $L_2$=0.08eV) is broader and more symmetric than in \327 ($L_1$=0.12 and $L_2$=0.05eV). It narrows down significantly in the metallic state, where a peak with $L$=0.025eV dominates. Although none of the spectra exhibits a well resolved peak-hump structure, as for the theoretical spectra, it may be present but hidden by broadening. Assuming this, the different lineshapes mean that the hump dominates in \214, the peak and the hump have similar contributions in \327 (building a more asymmetric shape) and the peak dominates in doped \327, as sketched in Fig. \ref{Fit}(c) for low temperatures. As there is no well defined separation between peak and hump, it is of course impossible to refine the fitting further. However, it gives a simple and natural way to explain the evolution from the more insulating state to the more metallic situation, by simply transferring weight from the broad hump to the peak. Such a decomposition was actually already proposed to describe the insulator to metal evolution of the lineshape in \327 doped with La \cite{HeNatMat15}. 

The very interesting point is now to use this underlying structure to better understand the evolution at $T_N$. The experimental high temperature spectra after background subtraction are compared in Fig. \ref{Fit}(b). The absence of sensitivity of \214 to $T_N$ can be explained if the incoherent hump is not sensitive to the magnetic order. This is in agreement with the theoretical scenario for the strongly correlated case, where most of the weight correspond to the Hubbard bands. In \327, most of the changes are due to the evolution of the coherent peak. More specifically, to explain a more symmetric lineshape, the coherent peak has to shift towards $E_F$, as sketched in the inset of Fig.\ref{Fit}(d). This is in agreement with the closure of the $\delta_S$ gap described before in the theory of the weakly correlated case. This suggests to identify the $\delta$ and $\delta_S$ energy scales. We note that the shift in \327 is only indirectly due to the closing of $\delta_S$. The main driving force for the shift is the relocation of the Fermi level at the middle of the $\Delta$ energy scale. Interestingly, there is a clear difference in theory for this position in the Mott and Slater cases at small dopings. In the Mott case, the QP peak forms at the edge of the Hubbard band without a gap. In contrast, in the Slater case, the QP forms at the edge of the remaining "Slater" gap $\delta_S$, around $\Delta$/2. In the experimental case, it seems that as long as a gap is open (either $\Delta$ or $\delta_S$), extrinsic degrees of freedom (disorder/impurities/dopant) may fix the position of the E$_F$ within the gap. Indeed at low temperatures, the peak poistions are similar in \214 and \327 for the same dopings. On the other hand, as soon as the low energy gap $\delta_S$ closes, E$_F$ is fixed to the theoretical position, inducing a shift in the weakly correlated case.

In Fig. \ref{Resistivity}, there is obviously a continuity between the metallicity found at high temperatures in the compounds keeping a magnetic ground state and the completely metallic ones. This decomposition bridges the two behaviors. The evolution in metallic \327 implies a rather large broadening of the peak with increasing temperature. Consequently, it is not well resolved from the hump anymore at high temperatures and the metallic nature of this spectrum is not obvious. Indeed, the lineshape for \327 pure and doped with 6\% La are very similar (Fig. \ref{Fit}b). This is why identifying the nature of the metallic state at high temperatures was difficult. This decomposition offers a qualitative understanding, but, here again, any more advanced fitting is impossible as it is difficult to separate not only the coherent part from incoherent part, but also the incoherent part from the background at high temperatures.

\vspace{0.2cm}

\section{Conclusion}
Iridates offer a unique opportunity to study the crossover from Mott to Slater behaviors as a function of correlation strength. Although there are many examples of Mott oxides with magnetic transitions \cite{KovalevaPRL04,GorelovPRL10}, there is often an orbital degree of freedom, which complicates the analysis of the mere impact of long-range magnetic order on the Mott state. This is not the case in iridates, where the filled \JJJ~states do not take an active part in the transition. Here, we have studied the evolution of ARPES lineshapes corresponding to the half-filled \J~across the temperature driven magnetic transition in \327 and \214 compounds, using different dopings to establish the universality of the behavior. We have then compared our results to theoretical expectations within the Dynamical Mean Field Theory. 

We argue that iridates are intermediately correlated systems, where the ARPES lineshape is formed by a continuity between incoherent Hubbard-like features and coherent QP-like excitations. In the more weakly correlated compound \327, a key change of the spectral lineshape at the magnetic transition is a broadening towards $E_F$, which we attribute to a shift of the coherent weight towards $E_F$. This result agrees with the closing of a coherent "Slater" gap at the magnetic transition expected by DMFT in the weakly correlated regime. This coherent weight coexists with incoherent features, centered at the position of Hubbard bands $\pm \Delta$/2. In contrast, we find that the behavior of \214 agrees more with the Hubbard-Mott scenario. The coherent part of the lineshape is small and no clear evolution is observed through \TN, as expected within the DMFT description. This decomposition of the spectra implies the existence of two energy scales, the large Mott-like gap $\Delta$, essentially fixed by short-range magnetic correlations and the small coherent gap $\delta_S$, fixed by the long range magnetic order. This is consistent with the absence of correlation between the large charge gap and the short-range magnetic order observed in \214 by spin-polarized STM \cite{ZhaoNatPhys19}.

We further find that in the weakly correlated case, the closing of the small magnetic gap $\delta_S$ redefines the position of the Fermi level in the middle of the Hubbard bands. This drives a shift of the whole spectrum towards this position, exactly starting and stopping over the width of the magnetic transition. This view is quite different from the Lifshitz-like transition that was proposed before, where the shift of the band would bring weight to the Fermi level \cite{SongMoonPRB18}. In future experimental and theoretical investigations, the role of disorder, also induced by the chemical substitution, should be investigated to fully describe the lineshapes and their behavior as a function of doping and temperature. The width of the magnetic transition, as well as the spectral changes observed by STM near defects and/or dopants \cite{OkadaNatMat13,SunPRR21,WangPNAS18}, implies that there is some degree of heterogeneity in these systems, which certainly plays a role in the absence of well resolved coherent peak in ARPES. 

\bibliography{Biblio_VsT}

\begin{thebibliography}{52}%
\makeatletter
\providecommand \@ifxundefined [1]{%
 \@ifx{#1\undefined}
}%
\providecommand \@ifnum [1]{%
 \ifnum #1\expandafter \@firstoftwo
 \else \expandafter \@secondoftwo
 \fi
}%
\providecommand \@ifx [1]{%
 \ifx #1\expandafter \@firstoftwo
 \else \expandafter \@secondoftwo
 \fi
}%
\providecommand \natexlab [1]{#1}%
\providecommand \enquote  [1]{``#1''}%
\providecommand \bibnamefont  [1]{#1}%
\providecommand \bibfnamefont [1]{#1}%
\providecommand \citenamefont [1]{#1}%
\providecommand \href@noop [0]{\@secondoftwo}%
\providecommand \href [0]{\begingroup \@sanitize@url \@href}%
\providecommand \@href[1]{\@@startlink{#1}\@@href}%
\providecommand \@@href[1]{\endgroup#1\@@endlink}%
\providecommand \@sanitize@url [0]{\catcode `\\12\catcode `\$12\catcode
  `\&12\catcode `\#12\catcode `\^12\catcode `\_12\catcode `\%12\relax}%
\providecommand \@@startlink[1]{}%
\providecommand \@@endlink[0]{}%
\providecommand \url  [0]{\begingroup\@sanitize@url \@url }%
\providecommand \@url [1]{\endgroup\@href {#1}{\urlprefix }}%
\providecommand \urlprefix  [0]{URL }%
\providecommand \Eprint [0]{\href }%
\providecommand \doibase [0]{http://dx.doi.org/}%
\providecommand \selectlanguage [0]{\@gobble}%
\providecommand \bibinfo  [0]{\@secondoftwo}%
\providecommand \bibfield  [0]{\@secondoftwo}%
\providecommand \translation [1]{[#1]}%
\providecommand \BibitemOpen [0]{}%
\providecommand \bibitemStop [0]{}%
\providecommand \bibitemNoStop [0]{.\EOS\space}%
\providecommand \EOS [0]{\spacefactor3000\relax}%
\providecommand \BibitemShut  [1]{\csname bibitem#1\endcsname}%
\let\auto@bib@innerbib\@empty
%</preamble>
\bibitem [{\citenamefont {Hao}\ \emph {et~al.}(2019)\citenamefont {Hao},
  \citenamefont {Wang}, \citenamefont {Yang}, \citenamefont {Meyers},
  \citenamefont {Sanchez}, \citenamefont {Fabbris}, \citenamefont {Choi},
  \citenamefont {Kim}, \citenamefont {Haskel}, \citenamefont {Ryan},
  \citenamefont {Barros}, \citenamefont {Chu}, \citenamefont {Dean},
  \citenamefont {Batista},\ and\ \citenamefont {Liu}}]{HaoNatCom19}%
  \BibitemOpen
  \bibfield  {author} {\bibinfo {author} {\bibfnamefont {L.}~\bibnamefont
  {Hao}}, \bibinfo {author} {\bibfnamefont {Z.}~\bibnamefont {Wang}}, \bibinfo
  {author} {\bibfnamefont {J.}~\bibnamefont {Yang}}, \bibinfo {author}
  {\bibfnamefont {D.}~\bibnamefont {Meyers}}, \bibinfo {author} {\bibfnamefont
  {J.}~\bibnamefont {Sanchez}}, \bibinfo {author} {\bibfnamefont
  {G.}~\bibnamefont {Fabbris}}, \bibinfo {author} {\bibfnamefont
  {Y.}~\bibnamefont {Choi}}, \bibinfo {author} {\bibfnamefont {J.-W.}\
  \bibnamefont {Kim}}, \bibinfo {author} {\bibfnamefont {D.}~\bibnamefont
  {Haskel}}, \bibinfo {author} {\bibfnamefont {P.~J.}\ \bibnamefont {Ryan}},
  \bibinfo {author} {\bibfnamefont {K.}~\bibnamefont {Barros}}, \bibinfo
  {author} {\bibfnamefont {J.-H.}\ \bibnamefont {Chu}}, \bibinfo {author}
  {\bibfnamefont {M.~P.~M.}\ \bibnamefont {Dean}}, \bibinfo {author}
  {\bibfnamefont {C.~D.}\ \bibnamefont {Batista}}, \ and\ \bibinfo {author}
  {\bibfnamefont {J.}~\bibnamefont {Liu}},\ }\href {\doibase
  10.1038/s41467-019-13271-6} {\bibfield  {journal} {\bibinfo  {journal}
  {Nature Communications}\ }\textbf {\bibinfo {volume} {10}},\ \bibinfo {pages}
  {5301} (\bibinfo {year} {2019})}\BibitemShut {NoStop}%
\bibitem [{\citenamefont {Fratino}\ \emph {et~al.}(2017)\citenamefont
  {Fratino}, \citenamefont {S\'emon}, \citenamefont {Charlebois}, \citenamefont
  {Sordi},\ and\ \citenamefont {Tremblay}}]{FratinoPRB17}%
  \BibitemOpen
  \bibfield  {author} {\bibinfo {author} {\bibfnamefont {L.}~\bibnamefont
  {Fratino}}, \bibinfo {author} {\bibfnamefont {P.}~\bibnamefont {S\'emon}},
  \bibinfo {author} {\bibfnamefont {M.}~\bibnamefont {Charlebois}}, \bibinfo
  {author} {\bibfnamefont {G.}~\bibnamefont {Sordi}}, \ and\ \bibinfo {author}
  {\bibfnamefont {A.-M.~S.}\ \bibnamefont {Tremblay}},\ }\href {\doibase
  10.1103/PhysRevB.95.235109} {\bibfield  {journal} {\bibinfo  {journal} {Phys.
  Rev. B}\ }\textbf {\bibinfo {volume} {95}},\ \bibinfo {pages} {235109}
  (\bibinfo {year} {2017})}\BibitemShut {NoStop}%
\bibitem [{\citenamefont {Camjayi}\ \emph {et~al.}(2006)\citenamefont
  {Camjayi}, \citenamefont {Chitra},\ and\ \citenamefont
  {Rozenberg}}]{CamjayiPRB06}%
  \BibitemOpen
  \bibfield  {author} {\bibinfo {author} {\bibfnamefont {A.}~\bibnamefont
  {Camjayi}}, \bibinfo {author} {\bibfnamefont {R.}~\bibnamefont {Chitra}}, \
  and\ \bibinfo {author} {\bibfnamefont {M.~J.}\ \bibnamefont {Rozenberg}},\
  }\href {\doibase 10.1103/PhysRevB.73.041103} {\bibfield  {journal} {\bibinfo
  {journal} {Phys. Rev. B}\ }\textbf {\bibinfo {volume} {73}},\ \bibinfo
  {pages} {041103} (\bibinfo {year} {2006})}\BibitemShut {NoStop}%
\bibitem [{\citenamefont {Kim}\ \emph {et~al.}(2008)\citenamefont {Kim},
  \citenamefont {Jin}, \citenamefont {Moon}, \citenamefont {Kim}, \citenamefont
  {Park}, \citenamefont {Leem}, \citenamefont {Yu}, \citenamefont {Noh},
  \citenamefont {Kim}, \citenamefont {Oh}, \citenamefont {Park}, \citenamefont
  {Durairaj}, \citenamefont {Cao},\ and\ \citenamefont
  {Rotenberg}}]{BJKimPRL08}%
  \BibitemOpen
  \bibfield  {author} {\bibinfo {author} {\bibfnamefont {B.~J.}\ \bibnamefont
  {Kim}}, \bibinfo {author} {\bibfnamefont {H.}~\bibnamefont {Jin}}, \bibinfo
  {author} {\bibfnamefont {S.~J.}\ \bibnamefont {Moon}}, \bibinfo {author}
  {\bibfnamefont {J.-Y.}\ \bibnamefont {Kim}}, \bibinfo {author} {\bibfnamefont
  {B.-G.}\ \bibnamefont {Park}}, \bibinfo {author} {\bibfnamefont {C.~S.}\
  \bibnamefont {Leem}}, \bibinfo {author} {\bibfnamefont {J.}~\bibnamefont
  {Yu}}, \bibinfo {author} {\bibfnamefont {T.~W.}\ \bibnamefont {Noh}},
  \bibinfo {author} {\bibfnamefont {C.}~\bibnamefont {Kim}}, \bibinfo {author}
  {\bibfnamefont {S.-J.}\ \bibnamefont {Oh}}, \bibinfo {author} {\bibfnamefont
  {J.-H.}\ \bibnamefont {Park}}, \bibinfo {author} {\bibfnamefont
  {V.}~\bibnamefont {Durairaj}}, \bibinfo {author} {\bibfnamefont
  {G.}~\bibnamefont {Cao}}, \ and\ \bibinfo {author} {\bibfnamefont
  {E.}~\bibnamefont {Rotenberg}},\ }\href@noop {} {\bibfield  {journal}
  {\bibinfo  {journal} {Phys. Rev. Lett.}\ }\textbf {\bibinfo {volume} {101}},\
  \bibinfo {pages} {076402} (\bibinfo {year} {2008})}\BibitemShut {NoStop}%
\bibitem [{\citenamefont {Martins}\ \emph {et~al.}(2018)\citenamefont
  {Martins}, \citenamefont {Lenz}, \citenamefont {Perfetti}, \citenamefont
  {Brouet}, \citenamefont {Bertran},\ and\ \citenamefont
  {Biermann}}]{MartinsPRM18}%
  \BibitemOpen
  \bibfield  {author} {\bibinfo {author} {\bibfnamefont {C.}~\bibnamefont
  {Martins}}, \bibinfo {author} {\bibfnamefont {B.}~\bibnamefont {Lenz}},
  \bibinfo {author} {\bibfnamefont {L.}~\bibnamefont {Perfetti}}, \bibinfo
  {author} {\bibfnamefont {V.}~\bibnamefont {Brouet}}, \bibinfo {author}
  {\bibfnamefont {F.}~\bibnamefont {Bertran}}, \ and\ \bibinfo {author}
  {\bibfnamefont {S.}~\bibnamefont {Biermann}},\ }\href {\doibase
  10.1103/PhysRevMaterials.2.032001} {\bibfield  {journal} {\bibinfo  {journal}
  {Phys. Rev. Materials}\ }\textbf {\bibinfo {volume} {2}},\ \bibinfo {pages}
  {032001} (\bibinfo {year} {2018})}\BibitemShut {NoStop}%
\bibitem [{\citenamefont {Moon}\ \emph {et~al.}(2008)\citenamefont {Moon},
  \citenamefont {Jin}, \citenamefont {Kim}, \citenamefont {Choi}, \citenamefont
  {Lee}, \citenamefont {Yu}, \citenamefont {Cao}, \citenamefont {Sumi},
  \citenamefont {Funakubo}, \citenamefont {Bernhard},\ and\ \citenamefont
  {Noh}}]{MoonPRL08}%
  \BibitemOpen
  \bibfield  {author} {\bibinfo {author} {\bibfnamefont {S.~J.}\ \bibnamefont
  {Moon}}, \bibinfo {author} {\bibfnamefont {H.}~\bibnamefont {Jin}}, \bibinfo
  {author} {\bibfnamefont {K.~W.}\ \bibnamefont {Kim}}, \bibinfo {author}
  {\bibfnamefont {W.~S.}\ \bibnamefont {Choi}}, \bibinfo {author}
  {\bibfnamefont {Y.~S.}\ \bibnamefont {Lee}}, \bibinfo {author} {\bibfnamefont
  {J.}~\bibnamefont {Yu}}, \bibinfo {author} {\bibfnamefont {G.}~\bibnamefont
  {Cao}}, \bibinfo {author} {\bibfnamefont {A.}~\bibnamefont {Sumi}}, \bibinfo
  {author} {\bibfnamefont {H.}~\bibnamefont {Funakubo}}, \bibinfo {author}
  {\bibfnamefont {C.}~\bibnamefont {Bernhard}}, \ and\ \bibinfo {author}
  {\bibfnamefont {T.~W.}\ \bibnamefont {Noh}},\ }\href {\doibase
  10.1103/PhysRevLett.101.226402} {\bibfield  {journal} {\bibinfo  {journal}
  {Phys. Rev. Lett.}\ }\textbf {\bibinfo {volume} {101}},\ \bibinfo {pages}
  {226402} (\bibinfo {year} {2008})}\BibitemShut {NoStop}%
\bibitem [{\citenamefont {Battisti}\ \emph {et~al.}(2017)\citenamefont
  {Battisti}, \citenamefont {Bastiaans}, \citenamefont {Fedoseev},
  \citenamefont {de~la Torre}, \citenamefont {Iliopoulos}, \citenamefont
  {Tamai}, \citenamefont {Hunter}, \citenamefont {Perry}, \citenamefont
  {Zaanen}, \citenamefont {Baumberger},\ and\ \citenamefont
  {Allan}}]{BattistiNatPhys17}%
  \BibitemOpen
  \bibfield  {author} {\bibinfo {author} {\bibfnamefont {I.}~\bibnamefont
  {Battisti}}, \bibinfo {author} {\bibfnamefont {K.~M.}\ \bibnamefont
  {Bastiaans}}, \bibinfo {author} {\bibfnamefont {V.}~\bibnamefont {Fedoseev}},
  \bibinfo {author} {\bibfnamefont {A.}~\bibnamefont {de~la Torre}}, \bibinfo
  {author} {\bibfnamefont {N.}~\bibnamefont {Iliopoulos}}, \bibinfo {author}
  {\bibfnamefont {A.}~\bibnamefont {Tamai}}, \bibinfo {author} {\bibfnamefont
  {E.~C.}\ \bibnamefont {Hunter}}, \bibinfo {author} {\bibfnamefont {R.~.~S.}\
  \bibnamefont {Perry}}, \bibinfo {author} {\bibfnamefont {J.}~\bibnamefont
  {Zaanen}}, \bibinfo {author} {\bibfnamefont {F.}~\bibnamefont {Baumberger}},
  \ and\ \bibinfo {author} {\bibfnamefont {M.~P.}\ \bibnamefont {Allan}},\
  }\href {\doibase 10.1038/nphys3894} {\bibfield  {journal} {\bibinfo
  {journal} {Nature Physics}\ }\textbf {\bibinfo {volume} {13}},\ \bibinfo
  {pages} {21} (\bibinfo {year} {2017})}\BibitemShut {NoStop}%
\bibitem [{\citenamefont {Zhao}\ \emph {et~al.}(2019)\citenamefont {Zhao},
  \citenamefont {Manna}, \citenamefont {Porter}, \citenamefont {Chen},
  \citenamefont {Uzdejczyk}, \citenamefont {Moodera}, \citenamefont {Wang},
  \citenamefont {Wilson},\ and\ \citenamefont {Zeljkovic}}]{ZhaoNatPhys19}%
  \BibitemOpen
  \bibfield  {author} {\bibinfo {author} {\bibfnamefont {H.}~\bibnamefont
  {Zhao}}, \bibinfo {author} {\bibfnamefont {S.}~\bibnamefont {Manna}},
  \bibinfo {author} {\bibfnamefont {Z.}~\bibnamefont {Porter}}, \bibinfo
  {author} {\bibfnamefont {X.}~\bibnamefont {Chen}}, \bibinfo {author}
  {\bibfnamefont {A.}~\bibnamefont {Uzdejczyk}}, \bibinfo {author}
  {\bibfnamefont {J.}~\bibnamefont {Moodera}}, \bibinfo {author} {\bibfnamefont
  {Z.}~\bibnamefont {Wang}}, \bibinfo {author} {\bibfnamefont {S.~D.}\
  \bibnamefont {Wilson}}, \ and\ \bibinfo {author} {\bibfnamefont
  {I.}~\bibnamefont {Zeljkovic}},\ }\href {\doibase 10.1038/s41567-019-0671-9}
  {\bibfield  {journal} {\bibinfo  {journal} {Nature Physics}\ }\textbf
  {\bibinfo {volume} {15}},\ \bibinfo {pages} {1267} (\bibinfo {year}
  {2019})}\BibitemShut {NoStop}%
\bibitem [{\citenamefont {Sun}\ \emph {et~al.}(2021)\citenamefont {Sun},
  \citenamefont {Guevara}, \citenamefont {Sykora}, \citenamefont {P\"arschke},
  \citenamefont {Manna}, \citenamefont {Maljuk}, \citenamefont {Wurmehl},
  \citenamefont {van~den Brink}, \citenamefont {B\"uchner},\ and\ \citenamefont
  {Hess}}]{SunPRR21}%
  \BibitemOpen
  \bibfield  {author} {\bibinfo {author} {\bibfnamefont {Z.}~\bibnamefont
  {Sun}}, \bibinfo {author} {\bibfnamefont {J.~M.}\ \bibnamefont {Guevara}},
  \bibinfo {author} {\bibfnamefont {S.}~\bibnamefont {Sykora}}, \bibinfo
  {author} {\bibfnamefont {E.~M.}\ \bibnamefont {P\"arschke}}, \bibinfo
  {author} {\bibfnamefont {K.}~\bibnamefont {Manna}}, \bibinfo {author}
  {\bibfnamefont {A.}~\bibnamefont {Maljuk}}, \bibinfo {author} {\bibfnamefont
  {S.}~\bibnamefont {Wurmehl}}, \bibinfo {author} {\bibfnamefont
  {J.}~\bibnamefont {van~den Brink}}, \bibinfo {author} {\bibfnamefont
  {B.}~\bibnamefont {B\"uchner}}, \ and\ \bibinfo {author} {\bibfnamefont
  {C.}~\bibnamefont {Hess}},\ }\href {\doibase
  10.1103/PhysRevResearch.3.023075} {\bibfield  {journal} {\bibinfo  {journal}
  {Phys. Rev. Research}\ }\textbf {\bibinfo {volume} {3}},\ \bibinfo {pages}
  {023075} (\bibinfo {year} {2021})}\BibitemShut {NoStop}%
\bibitem [{\citenamefont {Brouet}\ \emph {et~al.}(2015)\citenamefont {Brouet},
  \citenamefont {Mansart}, \citenamefont {Perfetti}, \citenamefont {Piovera},
  \citenamefont {Vobornik}, \citenamefont {{Le F{\`{e}}vre}}, \citenamefont
  {Bertran}, \citenamefont {Riggs}, \citenamefont {Shapiro}, \citenamefont
  {Giraldo-Gallo},\ and\ \citenamefont {Fisher}}]{BrouetPRB15}%
  \BibitemOpen
  \bibfield  {author} {\bibinfo {author} {\bibfnamefont {V.}~\bibnamefont
  {Brouet}}, \bibinfo {author} {\bibfnamefont {J.}~\bibnamefont {Mansart}},
  \bibinfo {author} {\bibfnamefont {L.}~\bibnamefont {Perfetti}}, \bibinfo
  {author} {\bibfnamefont {C.}~\bibnamefont {Piovera}}, \bibinfo {author}
  {\bibfnamefont {I.}~\bibnamefont {Vobornik}}, \bibinfo {author}
  {\bibfnamefont {P.}~\bibnamefont {{Le F{\`{e}}vre}}}, \bibinfo {author}
  {\bibfnamefont {F.}~\bibnamefont {Bertran}}, \bibinfo {author} {\bibfnamefont
  {S.~C.}\ \bibnamefont {Riggs}}, \bibinfo {author} {\bibfnamefont {M.~C.}\
  \bibnamefont {Shapiro}}, \bibinfo {author} {\bibfnamefont {P.}~\bibnamefont
  {Giraldo-Gallo}}, \ and\ \bibinfo {author} {\bibfnamefont {I.~R.}\
  \bibnamefont {Fisher}},\ }\href@noop {} {\bibfield  {journal} {\bibinfo
  {journal} {Physical Review B}\ }\textbf {\bibinfo {volume} {92}},\ \bibinfo
  {pages} {081117} (\bibinfo {year} {2015})}\BibitemShut {NoStop}%
\bibitem [{\citenamefont {Okada}\ \emph {et~al.}(2013)\citenamefont {Okada},
  \citenamefont {Walkup}, \citenamefont {Lin}, \citenamefont {Dhital},
  \citenamefont {Chang}, \citenamefont {Khadka}, \citenamefont {Zhou},
  \citenamefont {Jeng}, \citenamefont {Paranjape}, \citenamefont {Bansil},
  \citenamefont {Wang}, \citenamefont {Wilson},\ and\ \citenamefont
  {Madhavan}}]{OkadaNatMat13}%
  \BibitemOpen
  \bibfield  {author} {\bibinfo {author} {\bibfnamefont {Y.}~\bibnamefont
  {Okada}}, \bibinfo {author} {\bibfnamefont {D.}~\bibnamefont {Walkup}},
  \bibinfo {author} {\bibfnamefont {H.}~\bibnamefont {Lin}}, \bibinfo {author}
  {\bibfnamefont {C.}~\bibnamefont {Dhital}}, \bibinfo {author} {\bibfnamefont
  {T.-R.}\ \bibnamefont {Chang}}, \bibinfo {author} {\bibfnamefont
  {S.}~\bibnamefont {Khadka}}, \bibinfo {author} {\bibfnamefont
  {W.}~\bibnamefont {Zhou}}, \bibinfo {author} {\bibfnamefont {H.-T.}\
  \bibnamefont {Jeng}}, \bibinfo {author} {\bibfnamefont {M.}~\bibnamefont
  {Paranjape}}, \bibinfo {author} {\bibfnamefont {A.}~\bibnamefont {Bansil}},
  \bibinfo {author} {\bibfnamefont {Z.}~\bibnamefont {Wang}}, \bibinfo {author}
  {\bibfnamefont {S.~D.}\ \bibnamefont {Wilson}}, \ and\ \bibinfo {author}
  {\bibfnamefont {V.}~\bibnamefont {Madhavan}},\ }\href {\doibase
  10.1038/nmat3653} {\bibfield  {journal} {\bibinfo  {journal} {Nature
  Materials}\ }\textbf {\bibinfo {volume} {12}},\ \bibinfo {pages} {707}
  (\bibinfo {year} {2013})}\BibitemShut {NoStop}%
\bibitem [{\citenamefont {Zhao}\ \emph {et~al.}(2021)\citenamefont {Zhao},
  \citenamefont {Porter}, \citenamefont {Chen}, \citenamefont {Wilson},
  \citenamefont {Wang},\ and\ \citenamefont {Zeljkovic}}]{ZhaoScienceAdv21}%
  \BibitemOpen
  \bibfield  {author} {\bibinfo {author} {\bibfnamefont {H.}~\bibnamefont
  {Zhao}}, \bibinfo {author} {\bibfnamefont {Z.}~\bibnamefont {Porter}},
  \bibinfo {author} {\bibfnamefont {X.}~\bibnamefont {Chen}}, \bibinfo {author}
  {\bibfnamefont {S.~D.}\ \bibnamefont {Wilson}}, \bibinfo {author}
  {\bibfnamefont {Z.}~\bibnamefont {Wang}}, \ and\ \bibinfo {author}
  {\bibfnamefont {I.}~\bibnamefont {Zeljkovic}},\ }\href {\doibase
  10.1126/sciadv.abi6468} {\bibfield  {journal} {\bibinfo  {journal} {Science
  Advances}\ }\textbf {\bibinfo {volume} {7}},\ \bibinfo {pages} {eabi6468}
  (\bibinfo {year} {2021})},\ \Eprint
  {http://arxiv.org/abs/https://www.science.org/doi/pdf/10.1126/sciadv.abi6468}
  {https://www.science.org/doi/pdf/10.1126/sciadv.abi6468} \BibitemShut
  {NoStop}%
\bibitem [{\citenamefont {Brouet}\ \emph {et~al.}(2018)\citenamefont {Brouet},
  \citenamefont {Serrier-Garcia}, \citenamefont {Louat}, \citenamefont
  {Fruchter}, \citenamefont {Bertran}, \citenamefont {Le~F\`evre},
  \citenamefont {Rault}, \citenamefont {Forget},\ and\ \citenamefont
  {Colson}}]{BrouetPRB18_327}%
  \BibitemOpen
  \bibfield  {author} {\bibinfo {author} {\bibfnamefont {V.}~\bibnamefont
  {Brouet}}, \bibinfo {author} {\bibfnamefont {L.}~\bibnamefont
  {Serrier-Garcia}}, \bibinfo {author} {\bibfnamefont {A.}~\bibnamefont
  {Louat}}, \bibinfo {author} {\bibfnamefont {L.}~\bibnamefont {Fruchter}},
  \bibinfo {author} {\bibfnamefont {F.}~\bibnamefont {Bertran}}, \bibinfo
  {author} {\bibfnamefont {P.}~\bibnamefont {Le~F\`evre}}, \bibinfo {author}
  {\bibfnamefont {J.}~\bibnamefont {Rault}}, \bibinfo {author} {\bibfnamefont
  {A.}~\bibnamefont {Forget}}, \ and\ \bibinfo {author} {\bibfnamefont
  {D.}~\bibnamefont {Colson}},\ }\href {\doibase 10.1103/PhysRevB.98.235101}
  {\bibfield  {journal} {\bibinfo  {journal} {Phys. Rev. B}\ }\textbf {\bibinfo
  {volume} {98}},\ \bibinfo {pages} {235101} (\bibinfo {year}
  {2018})}\BibitemShut {NoStop}%
\bibitem [{\citenamefont {de~la Torre}\ \emph {et~al.}(2014)\citenamefont
  {de~la Torre}, \citenamefont {Hunter}, \citenamefont {Subedi}, \citenamefont
  {McKeown~Walker}, \citenamefont {Tamai}, \citenamefont {Kim}, \citenamefont
  {Hoesch}, \citenamefont {Perry}, \citenamefont {Georges},\ and\ \citenamefont
  {Baumberger}}]{DeLaTorrePRL14}%
  \BibitemOpen
  \bibfield  {author} {\bibinfo {author} {\bibfnamefont {A.}~\bibnamefont
  {de~la Torre}}, \bibinfo {author} {\bibfnamefont {E.}~\bibnamefont {Hunter}},
  \bibinfo {author} {\bibfnamefont {A.}~\bibnamefont {Subedi}}, \bibinfo
  {author} {\bibfnamefont {S.}~\bibnamefont {McKeown~Walker}}, \bibinfo
  {author} {\bibfnamefont {A.}~\bibnamefont {Tamai}}, \bibinfo {author}
  {\bibfnamefont {T.}~\bibnamefont {Kim}}, \bibinfo {author} {\bibfnamefont
  {M.}~\bibnamefont {Hoesch}}, \bibinfo {author} {\bibfnamefont
  {R.}~\bibnamefont {Perry}}, \bibinfo {author} {\bibfnamefont
  {A.}~\bibnamefont {Georges}}, \ and\ \bibinfo {author} {\bibfnamefont
  {F.}~\bibnamefont {Baumberger}},\ }\href@noop {} {\bibfield  {journal}
  {\bibinfo  {journal} {Phys. Rev. Lett.}\ }\textbf {\bibinfo {volume} {113}},\
  \bibinfo {pages} {256402} (\bibinfo {year} {2014})}\BibitemShut {NoStop}%
\bibitem [{sup()}]{sup}%
  \BibitemOpen
  \href@noop {} {\bibinfo  {journal} {See supplementary information}\
  }\BibitemShut {NoStop}%
\bibitem [{\citenamefont {Kim}\ \emph {et~al.}(2012{\natexlab{a}})\citenamefont
  {Kim}, \citenamefont {Casa}, \citenamefont {Upton}, \citenamefont {Gog},
  \citenamefont {Kim}, \citenamefont {Mitchell}, \citenamefont {van
  Veenendaal}, \citenamefont {Daghofer}, \citenamefont {van~den Brink},
  \citenamefont {Khaliullin},\ and\ \citenamefont {Kim}}]{KimPRL12_214}%
  \BibitemOpen
\bibfield  {journal} {  }\bibfield  {author} {\bibinfo {author} {\bibfnamefont
  {J.}~\bibnamefont {Kim}}, \bibinfo {author} {\bibfnamefont {D.}~\bibnamefont
  {Casa}}, \bibinfo {author} {\bibfnamefont {M.~H.}\ \bibnamefont {Upton}},
  \bibinfo {author} {\bibfnamefont {T.}~\bibnamefont {Gog}}, \bibinfo {author}
  {\bibfnamefont {Y.-J.}\ \bibnamefont {Kim}}, \bibinfo {author} {\bibfnamefont
  {J.~F.}\ \bibnamefont {Mitchell}}, \bibinfo {author} {\bibfnamefont
  {M.}~\bibnamefont {van Veenendaal}}, \bibinfo {author} {\bibfnamefont
  {M.}~\bibnamefont {Daghofer}}, \bibinfo {author} {\bibfnamefont
  {J.}~\bibnamefont {van~den Brink}}, \bibinfo {author} {\bibfnamefont
  {G.}~\bibnamefont {Khaliullin}}, \ and\ \bibinfo {author} {\bibfnamefont
  {B.~J.}\ \bibnamefont {Kim}},\ }\href {\doibase
  10.1103/PhysRevLett.108.177003} {\bibfield  {journal} {\bibinfo  {journal}
  {Phys. Rev. Lett.}\ }\textbf {\bibinfo {volume} {108}},\ \bibinfo {pages}
  {177003} (\bibinfo {year} {2012}{\natexlab{a}})}\BibitemShut {NoStop}%
\bibitem [{\citenamefont {Dhital}\ \emph {et~al.}(2013)\citenamefont {Dhital},
  \citenamefont {Hogan}, \citenamefont {Yamani}, \citenamefont {de~la Cruz},
  \citenamefont {Chen}, \citenamefont {Khadka}, \citenamefont {Ren},\ and\
  \citenamefont {Wilson}}]{DhitalPRB13}%
  \BibitemOpen
  \bibfield  {author} {\bibinfo {author} {\bibfnamefont {C.}~\bibnamefont
  {Dhital}}, \bibinfo {author} {\bibfnamefont {T.}~\bibnamefont {Hogan}},
  \bibinfo {author} {\bibfnamefont {Z.}~\bibnamefont {Yamani}}, \bibinfo
  {author} {\bibfnamefont {C.}~\bibnamefont {de~la Cruz}}, \bibinfo {author}
  {\bibfnamefont {X.}~\bibnamefont {Chen}}, \bibinfo {author} {\bibfnamefont
  {S.}~\bibnamefont {Khadka}}, \bibinfo {author} {\bibfnamefont
  {Z.}~\bibnamefont {Ren}}, \ and\ \bibinfo {author} {\bibfnamefont {S.~D.}\
  \bibnamefont {Wilson}},\ }\href {\doibase 10.1103/PhysRevB.87.144405}
  {\bibfield  {journal} {\bibinfo  {journal} {Phys. Rev. B}\ }\textbf {\bibinfo
  {volume} {87}},\ \bibinfo {pages} {144405} (\bibinfo {year}
  {2013})}\BibitemShut {NoStop}%
\bibitem [{\citenamefont {Kim}\ \emph {et~al.}(2012{\natexlab{b}})\citenamefont
  {Kim}, \citenamefont {Said}, \citenamefont {Casa}, \citenamefont {Upton},
  \citenamefont {Gog}, \citenamefont {Daghofer}, \citenamefont {Jackeli},
  \citenamefont {van~den Brink}, \citenamefont {Khaliullin},\ and\
  \citenamefont {Kim}}]{KimPRL12_327}%
  \BibitemOpen
  \bibfield  {author} {\bibinfo {author} {\bibfnamefont {J.}~\bibnamefont
  {Kim}}, \bibinfo {author} {\bibfnamefont {A.~H.}\ \bibnamefont {Said}},
  \bibinfo {author} {\bibfnamefont {D.}~\bibnamefont {Casa}}, \bibinfo {author}
  {\bibfnamefont {M.~H.}\ \bibnamefont {Upton}}, \bibinfo {author}
  {\bibfnamefont {T.}~\bibnamefont {Gog}}, \bibinfo {author} {\bibfnamefont
  {M.}~\bibnamefont {Daghofer}}, \bibinfo {author} {\bibfnamefont
  {G.}~\bibnamefont {Jackeli}}, \bibinfo {author} {\bibfnamefont
  {J.}~\bibnamefont {van~den Brink}}, \bibinfo {author} {\bibfnamefont
  {G.}~\bibnamefont {Khaliullin}}, \ and\ \bibinfo {author} {\bibfnamefont
  {B.~J.}\ \bibnamefont {Kim}},\ }\href {\doibase
  10.1103/PhysRevLett.109.157402} {\bibfield  {journal} {\bibinfo  {journal}
  {Phys. Rev. Lett.}\ }\textbf {\bibinfo {volume} {109}},\ \bibinfo {pages}
  {157402} (\bibinfo {year} {2012}{\natexlab{b}})}\BibitemShut {NoStop}%
\bibitem [{\citenamefont {Korneta}\ \emph {et~al.}(2010)\citenamefont
  {Korneta}, \citenamefont {Qi}, \citenamefont {Chikara}, \citenamefont
  {Parkin}, \citenamefont {De~Long}, \citenamefont {Schlottmann},\ and\
  \citenamefont {Cao}}]{KornetaPRB10}%
  \BibitemOpen
  \bibfield  {author} {\bibinfo {author} {\bibfnamefont {O.~B.}\ \bibnamefont
  {Korneta}}, \bibinfo {author} {\bibfnamefont {T.}~\bibnamefont {Qi}},
  \bibinfo {author} {\bibfnamefont {S.}~\bibnamefont {Chikara}}, \bibinfo
  {author} {\bibfnamefont {S.}~\bibnamefont {Parkin}}, \bibinfo {author}
  {\bibfnamefont {L.~E.}\ \bibnamefont {De~Long}}, \bibinfo {author}
  {\bibfnamefont {P.}~\bibnamefont {Schlottmann}}, \ and\ \bibinfo {author}
  {\bibfnamefont {G.}~\bibnamefont {Cao}},\ }\href@noop {} {\bibfield
  {journal} {\bibinfo  {journal} {Phys. Rev. B}\ }\textbf {\bibinfo {volume}
  {82}},\ \bibinfo {pages} {115117} (\bibinfo {year} {2010})}\BibitemShut
  {NoStop}%
\bibitem [{\citenamefont {Cao}\ \emph {et~al.}(2002)\citenamefont {Cao},
  \citenamefont {Xin}, \citenamefont {Alexander}, \citenamefont {Crow},
  \citenamefont {Schlottmann}, \citenamefont {Crawford}, \citenamefont
  {Harlow},\ and\ \citenamefont {Marshall}}]{CaoPRB02}%
  \BibitemOpen
  \bibfield  {author} {\bibinfo {author} {\bibfnamefont {G.}~\bibnamefont
  {Cao}}, \bibinfo {author} {\bibfnamefont {Y.}~\bibnamefont {Xin}}, \bibinfo
  {author} {\bibfnamefont {C.~S.}\ \bibnamefont {Alexander}}, \bibinfo {author}
  {\bibfnamefont {J.~E.}\ \bibnamefont {Crow}}, \bibinfo {author}
  {\bibfnamefont {P.}~\bibnamefont {Schlottmann}}, \bibinfo {author}
  {\bibfnamefont {M.~K.}\ \bibnamefont {Crawford}}, \bibinfo {author}
  {\bibfnamefont {R.~L.}\ \bibnamefont {Harlow}}, \ and\ \bibinfo {author}
  {\bibfnamefont {W.}~\bibnamefont {Marshall}},\ }\href {\doibase
  10.1103/PhysRevB.66.214412} {\bibfield  {journal} {\bibinfo  {journal} {Phys.
  Rev. B}\ }\textbf {\bibinfo {volume} {66}},\ \bibinfo {pages} {214412}
  (\bibinfo {year} {2002})}\BibitemShut {NoStop}%
\bibitem [{\citenamefont {Chen}\ \emph {et~al.}(2015)\citenamefont {Chen},
  \citenamefont {Hogan}, \citenamefont {Walkup}, \citenamefont {Zhou},
  \citenamefont {Pokharel}, \citenamefont {Yao}, \citenamefont {Tian},
  \citenamefont {Ward}, \citenamefont {Zhao}, \citenamefont {Parshall},
  \citenamefont {Opeil}, \citenamefont {Lynn}, \citenamefont {Madhavan},\ and\
  \citenamefont {Wilson}}]{ChenPRB15}%
  \BibitemOpen
  \bibfield  {author} {\bibinfo {author} {\bibfnamefont {X.}~\bibnamefont
  {Chen}}, \bibinfo {author} {\bibfnamefont {T.}~\bibnamefont {Hogan}},
  \bibinfo {author} {\bibfnamefont {D.}~\bibnamefont {Walkup}}, \bibinfo
  {author} {\bibfnamefont {W.}~\bibnamefont {Zhou}}, \bibinfo {author}
  {\bibfnamefont {M.}~\bibnamefont {Pokharel}}, \bibinfo {author}
  {\bibfnamefont {M.}~\bibnamefont {Yao}}, \bibinfo {author} {\bibfnamefont
  {W.}~\bibnamefont {Tian}}, \bibinfo {author} {\bibfnamefont {T.~Z.}\
  \bibnamefont {Ward}}, \bibinfo {author} {\bibfnamefont {Y.}~\bibnamefont
  {Zhao}}, \bibinfo {author} {\bibfnamefont {D.}~\bibnamefont {Parshall}},
  \bibinfo {author} {\bibfnamefont {C.}~\bibnamefont {Opeil}}, \bibinfo
  {author} {\bibfnamefont {J.~W.}\ \bibnamefont {Lynn}}, \bibinfo {author}
  {\bibfnamefont {V.}~\bibnamefont {Madhavan}}, \ and\ \bibinfo {author}
  {\bibfnamefont {S.~D.}\ \bibnamefont {Wilson}},\ }\href {\doibase
  10.1103/PhysRevB.92.075125} {\bibfield  {journal} {\bibinfo  {journal} {Phys.
  Rev. B}\ }\textbf {\bibinfo {volume} {92}},\ \bibinfo {pages} {075125}
  (\bibinfo {year} {2015})}\BibitemShut {NoStop}%
\bibitem [{\citenamefont {Moon}\ \emph {et~al.}(2009)\citenamefont {Moon},
  \citenamefont {Jin}, \citenamefont {Choi}, \citenamefont {Lee}, \citenamefont
  {Seo}, \citenamefont {Yu}, \citenamefont {Cao}, \citenamefont {Noh},\ and\
  \citenamefont {Lee}}]{MoonPRB09}%
  \BibitemOpen
  \bibfield  {author} {\bibinfo {author} {\bibfnamefont {S.}~\bibnamefont
  {Moon}}, \bibinfo {author} {\bibfnamefont {H.}~\bibnamefont {Jin}}, \bibinfo
  {author} {\bibfnamefont {W.}~\bibnamefont {Choi}}, \bibinfo {author}
  {\bibfnamefont {J.}~\bibnamefont {Lee}}, \bibinfo {author} {\bibfnamefont
  {S.}~\bibnamefont {Seo}}, \bibinfo {author} {\bibfnamefont {J.}~\bibnamefont
  {Yu}}, \bibinfo {author} {\bibfnamefont {G.}~\bibnamefont {Cao}}, \bibinfo
  {author} {\bibfnamefont {T.}~\bibnamefont {Noh}}, \ and\ \bibinfo {author}
  {\bibfnamefont {Y.}~\bibnamefont {Lee}},\ }\href@noop {} {\bibfield
  {journal} {\bibinfo  {journal} {Phys. Rev. B}\ }\textbf {\bibinfo {volume}
  {80}},\ \bibinfo {pages} {195110} (\bibinfo {year} {2009})}\BibitemShut
  {NoStop}%
\bibitem [{\citenamefont {Ahn}\ \emph {et~al.}(2016)\citenamefont {Ahn},
  \citenamefont {Song}, \citenamefont {Hogan}, \citenamefont {Wilson},\ and\
  \citenamefont {Moon}}]{AhnSciRep16}%
  \BibitemOpen
  \bibfield  {author} {\bibinfo {author} {\bibfnamefont {G.}~\bibnamefont
  {Ahn}}, \bibinfo {author} {\bibfnamefont {S.}~\bibnamefont {Song}}, \bibinfo
  {author} {\bibfnamefont {T.}~\bibnamefont {Hogan}}, \bibinfo {author}
  {\bibfnamefont {S.~D.}\ \bibnamefont {Wilson}}, \ and\ \bibinfo {author}
  {\bibfnamefont {S.}~\bibnamefont {Moon}},\ }\href@noop {} {\bibfield
  {journal} {\bibinfo  {journal} {Scientific reports}\ }\textbf {\bibinfo
  {volume} {6}},\ \bibinfo {pages} {32632} (\bibinfo {year}
  {2016})}\BibitemShut {NoStop}%
\bibitem [{\citenamefont {Xu}\ \emph {et~al.}(2020)\citenamefont {Xu},
  \citenamefont {Marsik}, \citenamefont {Sheveleva}, \citenamefont {Lyzwa},
  \citenamefont {Louat}, \citenamefont {Brouet}, \citenamefont {Munzar},\ and\
  \citenamefont {Bernhard}}]{XuBernhardPRL20}%
  \BibitemOpen
  \bibfield  {author} {\bibinfo {author} {\bibfnamefont {B.}~\bibnamefont
  {Xu}}, \bibinfo {author} {\bibfnamefont {P.}~\bibnamefont {Marsik}}, \bibinfo
  {author} {\bibfnamefont {E.}~\bibnamefont {Sheveleva}}, \bibinfo {author}
  {\bibfnamefont {F.}~\bibnamefont {Lyzwa}}, \bibinfo {author} {\bibfnamefont
  {A.}~\bibnamefont {Louat}}, \bibinfo {author} {\bibfnamefont
  {V.}~\bibnamefont {Brouet}}, \bibinfo {author} {\bibfnamefont
  {D.}~\bibnamefont {Munzar}}, \ and\ \bibinfo {author} {\bibfnamefont
  {C.}~\bibnamefont {Bernhard}},\ }\href {\doibase
  10.1103/PhysRevLett.124.027402} {\bibfield  {journal} {\bibinfo  {journal}
  {Phys. Rev. Lett.}\ }\textbf {\bibinfo {volume} {124}},\ \bibinfo {pages}
  {027402} (\bibinfo {year} {2020})}\BibitemShut {NoStop}%
\bibitem [{\citenamefont {Sohn}\ \emph {et~al.}(2014)\citenamefont {Sohn},
  \citenamefont {Lee}, \citenamefont {Park}, \citenamefont {Noh}, \citenamefont
  {Yoo}, \citenamefont {Moon}, \citenamefont {Kim}, \citenamefont {Qi},
  \citenamefont {Cao}, \citenamefont {Cho},\ and\ \citenamefont
  {Noh}}]{SohnPRB14}%
  \BibitemOpen
  \bibfield  {author} {\bibinfo {author} {\bibfnamefont {C.~H.}\ \bibnamefont
  {Sohn}}, \bibinfo {author} {\bibfnamefont {M.-C.}\ \bibnamefont {Lee}},
  \bibinfo {author} {\bibfnamefont {H.~J.}\ \bibnamefont {Park}}, \bibinfo
  {author} {\bibfnamefont {K.~J.}\ \bibnamefont {Noh}}, \bibinfo {author}
  {\bibfnamefont {H.~K.}\ \bibnamefont {Yoo}}, \bibinfo {author} {\bibfnamefont
  {S.~J.}\ \bibnamefont {Moon}}, \bibinfo {author} {\bibfnamefont {K.~W.}\
  \bibnamefont {Kim}}, \bibinfo {author} {\bibfnamefont {T.~F.}\ \bibnamefont
  {Qi}}, \bibinfo {author} {\bibfnamefont {G.}~\bibnamefont {Cao}}, \bibinfo
  {author} {\bibfnamefont {D.-Y.}\ \bibnamefont {Cho}}, \ and\ \bibinfo
  {author} {\bibfnamefont {T.~W.}\ \bibnamefont {Noh}},\ }\href {\doibase
  10.1103/PhysRevB.90.041105} {\bibfield  {journal} {\bibinfo  {journal} {Phys.
  Rev. B}\ }\textbf {\bibinfo {volume} {90}},\ \bibinfo {pages} {041105}
  (\bibinfo {year} {2014})}\BibitemShut {NoStop}%
\bibitem [{\citenamefont {Song}\ \emph {et~al.}(2018)\citenamefont {Song},
  \citenamefont {Kim}, \citenamefont {Ahn}, \citenamefont {Seo}, \citenamefont
  {Schmehr}, \citenamefont {Aling}, \citenamefont {Wilson}, \citenamefont
  {Kim},\ and\ \citenamefont {Moon}}]{SongMoonPRB18}%
  \BibitemOpen
  \bibfield  {author} {\bibinfo {author} {\bibfnamefont {S.}~\bibnamefont
  {Song}}, \bibinfo {author} {\bibfnamefont {S.}~\bibnamefont {Kim}}, \bibinfo
  {author} {\bibfnamefont {G.~H.}\ \bibnamefont {Ahn}}, \bibinfo {author}
  {\bibfnamefont {J.~H.}\ \bibnamefont {Seo}}, \bibinfo {author} {\bibfnamefont
  {J.~L.}\ \bibnamefont {Schmehr}}, \bibinfo {author} {\bibfnamefont
  {M.}~\bibnamefont {Aling}}, \bibinfo {author} {\bibfnamefont {S.~D.}\
  \bibnamefont {Wilson}}, \bibinfo {author} {\bibfnamefont {Y.~K.}\
  \bibnamefont {Kim}}, \ and\ \bibinfo {author} {\bibfnamefont {S.~J.}\
  \bibnamefont {Moon}},\ }\href {\doibase 10.1103/PhysRevB.98.035110}
  {\bibfield  {journal} {\bibinfo  {journal} {Phys. Rev. B}\ }\textbf {\bibinfo
  {volume} {98}},\ \bibinfo {pages} {035110} (\bibinfo {year}
  {2018})}\BibitemShut {NoStop}%
\bibitem [{\citenamefont {Wang}\ \emph {et~al.}(2019)\citenamefont {Wang},
  \citenamefont {Walkup}, \citenamefont {Maximenko}, \citenamefont {Zhou},
  \citenamefont {Hogan}, \citenamefont {Wang}, \citenamefont {Wilson},\ and\
  \citenamefont {Madhavan}}]{WangNPJquantum19}%
  \BibitemOpen
  \bibfield  {author} {\bibinfo {author} {\bibfnamefont {Z.}~\bibnamefont
  {Wang}}, \bibinfo {author} {\bibfnamefont {D.}~\bibnamefont {Walkup}},
  \bibinfo {author} {\bibfnamefont {Y.}~\bibnamefont {Maximenko}}, \bibinfo
  {author} {\bibfnamefont {W.}~\bibnamefont {Zhou}}, \bibinfo {author}
  {\bibfnamefont {T.}~\bibnamefont {Hogan}}, \bibinfo {author} {\bibfnamefont
  {Z.}~\bibnamefont {Wang}}, \bibinfo {author} {\bibfnamefont {S.~D.}\
  \bibnamefont {Wilson}}, \ and\ \bibinfo {author} {\bibfnamefont
  {V.}~\bibnamefont {Madhavan}},\ }\href {\doibase 10.1038/s41535-019-0183-y}
  {\bibfield  {journal} {\bibinfo  {journal} {npj Quantum Materials}\ }\textbf
  {\bibinfo {volume} {4}},\ \bibinfo {pages} {43} (\bibinfo {year}
  {2019})}\BibitemShut {NoStop}%
\bibitem [{\citenamefont {King}\ \emph {et~al.}(2013)\citenamefont {King},
  \citenamefont {Takayama}, \citenamefont {Tamai}, \citenamefont {Rozbicki},
  \citenamefont {Walker}, \citenamefont {Shi}, \citenamefont {Patthey},
  \citenamefont {Moore}, \citenamefont {Lu}, \citenamefont {Shen},
  \citenamefont {Takagi},\ and\ \citenamefont {Baumberger}}]{KingPRB13}%
  \BibitemOpen
  \bibfield  {author} {\bibinfo {author} {\bibfnamefont {P.~D.~C.}\
  \bibnamefont {King}}, \bibinfo {author} {\bibfnamefont {T.}~\bibnamefont
  {Takayama}}, \bibinfo {author} {\bibfnamefont {A.}~\bibnamefont {Tamai}},
  \bibinfo {author} {\bibfnamefont {E.}~\bibnamefont {Rozbicki}}, \bibinfo
  {author} {\bibfnamefont {S.~M.}\ \bibnamefont {Walker}}, \bibinfo {author}
  {\bibfnamefont {M.}~\bibnamefont {Shi}}, \bibinfo {author} {\bibfnamefont
  {L.}~\bibnamefont {Patthey}}, \bibinfo {author} {\bibfnamefont {R.~G.}\
  \bibnamefont {Moore}}, \bibinfo {author} {\bibfnamefont {D.}~\bibnamefont
  {Lu}}, \bibinfo {author} {\bibfnamefont {K.~M.}\ \bibnamefont {Shen}},
  \bibinfo {author} {\bibfnamefont {H.}~\bibnamefont {Takagi}}, \ and\ \bibinfo
  {author} {\bibfnamefont {F.}~\bibnamefont {Baumberger}},\ }\href {\doibase
  10.1103/PhysRevB.87.241106} {\bibfield  {journal} {\bibinfo  {journal} {Phys.
  Rev. B}\ }\textbf {\bibinfo {volume} {87}},\ \bibinfo {pages} {241106}
  (\bibinfo {year} {2013})}\BibitemShut {NoStop}%
\bibitem [{\citenamefont {Affeldt}\ \emph {et~al.}(2017)\citenamefont
  {Affeldt}, \citenamefont {Hogan}, \citenamefont {Smallwood}, \citenamefont
  {Das}, \citenamefont {Denlinger}, \citenamefont {Wilson}, \citenamefont
  {Vishwanath},\ and\ \citenamefont {Lanzara}}]{AffeldtPRB17_VsT}%
  \BibitemOpen
  \bibfield  {author} {\bibinfo {author} {\bibfnamefont {G.}~\bibnamefont
  {Affeldt}}, \bibinfo {author} {\bibfnamefont {T.}~\bibnamefont {Hogan}},
  \bibinfo {author} {\bibfnamefont {C.~L.}\ \bibnamefont {Smallwood}}, \bibinfo
  {author} {\bibfnamefont {T.}~\bibnamefont {Das}}, \bibinfo {author}
  {\bibfnamefont {J.~D.}\ \bibnamefont {Denlinger}}, \bibinfo {author}
  {\bibfnamefont {S.~D.}\ \bibnamefont {Wilson}}, \bibinfo {author}
  {\bibfnamefont {A.}~\bibnamefont {Vishwanath}}, \ and\ \bibinfo {author}
  {\bibfnamefont {A.}~\bibnamefont {Lanzara}},\ }\href {\doibase
  10.1103/PhysRevB.95.235151} {\bibfield  {journal} {\bibinfo  {journal} {Phys.
  Rev. B}\ }\textbf {\bibinfo {volume} {95}},\ \bibinfo {pages} {235151}
  (\bibinfo {year} {2017})}\BibitemShut {NoStop}%
\bibitem [{\citenamefont {Ge}\ \emph {et~al.}(2011)\citenamefont {Ge},
  \citenamefont {Qi}, \citenamefont {Korneta}, \citenamefont {De~Long},
  \citenamefont {Schlottmann}, \citenamefont {Crummett},\ and\ \citenamefont
  {Cao}}]{GeCaoPRB11}%
  \BibitemOpen
  \bibfield  {author} {\bibinfo {author} {\bibfnamefont {M.}~\bibnamefont
  {Ge}}, \bibinfo {author} {\bibfnamefont {T.}~\bibnamefont {Qi}}, \bibinfo
  {author} {\bibfnamefont {O.}~\bibnamefont {Korneta}}, \bibinfo {author}
  {\bibfnamefont {D.}~\bibnamefont {De~Long}}, \bibinfo {author} {\bibfnamefont
  {P.}~\bibnamefont {Schlottmann}}, \bibinfo {author} {\bibfnamefont
  {W.}~\bibnamefont {Crummett}}, \ and\ \bibinfo {author} {\bibfnamefont
  {G.}~\bibnamefont {Cao}},\ }\href@noop {} {\bibfield  {journal} {\bibinfo
  {journal} {Phys. Rev. B}\ }\textbf {\bibinfo {volume} {84}},\ \bibinfo
  {pages} {100402} (\bibinfo {year} {2011})}\BibitemShut {NoStop}%
\bibitem [{\citenamefont {Hogan}\ \emph {et~al.}(2015)\citenamefont {Hogan},
  \citenamefont {Yamani}, \citenamefont {Walkup}, \citenamefont {Chen},
  \citenamefont {Dally}, \citenamefont {Ward}, \citenamefont {Dean},
  \citenamefont {Hill}, \citenamefont {Islam}, \citenamefont {Madhavan},\ and\
  \citenamefont {Wilson}}]{HoganPRL15}%
  \BibitemOpen
  \bibfield  {author} {\bibinfo {author} {\bibfnamefont {T.}~\bibnamefont
  {Hogan}}, \bibinfo {author} {\bibfnamefont {Z.}~\bibnamefont {Yamani}},
  \bibinfo {author} {\bibfnamefont {D.}~\bibnamefont {Walkup}}, \bibinfo
  {author} {\bibfnamefont {X.}~\bibnamefont {Chen}}, \bibinfo {author}
  {\bibfnamefont {R.}~\bibnamefont {Dally}}, \bibinfo {author} {\bibfnamefont
  {T.~Z.}\ \bibnamefont {Ward}}, \bibinfo {author} {\bibfnamefont {M.~P.~M.}\
  \bibnamefont {Dean}}, \bibinfo {author} {\bibfnamefont {J.}~\bibnamefont
  {Hill}}, \bibinfo {author} {\bibfnamefont {Z.}~\bibnamefont {Islam}},
  \bibinfo {author} {\bibfnamefont {V.}~\bibnamefont {Madhavan}}, \ and\
  \bibinfo {author} {\bibfnamefont {S.~D.}\ \bibnamefont {Wilson}},\ }\href
  {\doibase 10.1103/PhysRevLett.114.257203} {\bibfield  {journal} {\bibinfo
  {journal} {Phys. Rev. Lett.}\ }\textbf {\bibinfo {volume} {114}},\ \bibinfo
  {pages} {257203} (\bibinfo {year} {2015})}\BibitemShut {NoStop}%
\bibitem [{\citenamefont {Dhital}\ \emph {et~al.}(2014)\citenamefont {Dhital},
  \citenamefont {Hogan}, \citenamefont {Zhou}, \citenamefont {Chen},
  \citenamefont {Ren}, \citenamefont {Pokharel}, \citenamefont {Okada},
  \citenamefont {Heine}, \citenamefont {Tian}, \citenamefont {Yamani},
  \citenamefont {Opeil}, \citenamefont {Helton}, \citenamefont {Lynn},
  \citenamefont {Wang}, \citenamefont {Madhavan},\ and\ \citenamefont
  {Wilson}}]{DhitalNatCom14}%
  \BibitemOpen
  \bibfield  {author} {\bibinfo {author} {\bibfnamefont {C.}~\bibnamefont
  {Dhital}}, \bibinfo {author} {\bibfnamefont {T.}~\bibnamefont {Hogan}},
  \bibinfo {author} {\bibfnamefont {W.}~\bibnamefont {Zhou}}, \bibinfo {author}
  {\bibfnamefont {X.}~\bibnamefont {Chen}}, \bibinfo {author} {\bibfnamefont
  {Z.}~\bibnamefont {Ren}}, \bibinfo {author} {\bibfnamefont {M.}~\bibnamefont
  {Pokharel}}, \bibinfo {author} {\bibfnamefont {Y.}~\bibnamefont {Okada}},
  \bibinfo {author} {\bibfnamefont {M.}~\bibnamefont {Heine}}, \bibinfo
  {author} {\bibfnamefont {W.}~\bibnamefont {Tian}}, \bibinfo {author}
  {\bibfnamefont {Z.}~\bibnamefont {Yamani}}, \bibinfo {author} {\bibfnamefont
  {C.}~\bibnamefont {Opeil}}, \bibinfo {author} {\bibfnamefont {J.~S.}\
  \bibnamefont {Helton}}, \bibinfo {author} {\bibfnamefont {J.~W.}\
  \bibnamefont {Lynn}}, \bibinfo {author} {\bibfnamefont {Z.}~\bibnamefont
  {Wang}}, \bibinfo {author} {\bibfnamefont {V.}~\bibnamefont {Madhavan}}, \
  and\ \bibinfo {author} {\bibfnamefont {S.~D.}\ \bibnamefont {Wilson}},\
  }\href {\doibase 10.1038/ncomms4377} {\bibfield  {journal} {\bibinfo
  {journal} {Nature Communications}\ }\textbf {\bibinfo {volume} {5}},\
  \bibinfo {pages} {3377} (\bibinfo {year} {2014})}\BibitemShut {NoStop}%
\bibitem [{\citenamefont {Yuan}\ \emph {et~al.}(2015)\citenamefont {Yuan},
  \citenamefont {Aswartham}, \citenamefont {Terzic}, \citenamefont {Zheng},
  \citenamefont {Zhao}, \citenamefont {Schlottmann},\ and\ \citenamefont
  {Cao}}]{YuanPRB15}%
  \BibitemOpen
  \bibfield  {author} {\bibinfo {author} {\bibfnamefont {S.~J.}\ \bibnamefont
  {Yuan}}, \bibinfo {author} {\bibfnamefont {S.}~\bibnamefont {Aswartham}},
  \bibinfo {author} {\bibfnamefont {J.}~\bibnamefont {Terzic}}, \bibinfo
  {author} {\bibfnamefont {H.}~\bibnamefont {Zheng}}, \bibinfo {author}
  {\bibfnamefont {H.~D.}\ \bibnamefont {Zhao}}, \bibinfo {author}
  {\bibfnamefont {P.}~\bibnamefont {Schlottmann}}, \ and\ \bibinfo {author}
  {\bibfnamefont {G.}~\bibnamefont {Cao}},\ }\href {\doibase
  10.1103/PhysRevB.92.245103} {\bibfield  {journal} {\bibinfo  {journal} {Phys.
  Rev. B}\ }\textbf {\bibinfo {volume} {92}},\ \bibinfo {pages} {245103}
  (\bibinfo {year} {2015})}\BibitemShut {NoStop}%
\bibitem [{\citenamefont {Calder}\ \emph {et~al.}(2015)\citenamefont {Calder},
  \citenamefont {Kim}, \citenamefont {Cao}, \citenamefont {Cantoni},
  \citenamefont {May}, \citenamefont {Cao}, \citenamefont {Aczel},
  \citenamefont {Matsuda}, \citenamefont {Choi}, \citenamefont {Haskel},
  \citenamefont {Sales}, \citenamefont {Mandrus}, \citenamefont {Lumsden},\
  and\ \citenamefont {Christianson}}]{CalderPRB15}%
  \BibitemOpen
  \bibfield  {author} {\bibinfo {author} {\bibfnamefont {S.}~\bibnamefont
  {Calder}}, \bibinfo {author} {\bibfnamefont {J.~W.}\ \bibnamefont {Kim}},
  \bibinfo {author} {\bibfnamefont {G.-X.}\ \bibnamefont {Cao}}, \bibinfo
  {author} {\bibfnamefont {C.}~\bibnamefont {Cantoni}}, \bibinfo {author}
  {\bibfnamefont {A.~F.}\ \bibnamefont {May}}, \bibinfo {author} {\bibfnamefont
  {H.~B.}\ \bibnamefont {Cao}}, \bibinfo {author} {\bibfnamefont {A.~A.}\
  \bibnamefont {Aczel}}, \bibinfo {author} {\bibfnamefont {M.}~\bibnamefont
  {Matsuda}}, \bibinfo {author} {\bibfnamefont {Y.}~\bibnamefont {Choi}},
  \bibinfo {author} {\bibfnamefont {D.}~\bibnamefont {Haskel}}, \bibinfo
  {author} {\bibfnamefont {B.~C.}\ \bibnamefont {Sales}}, \bibinfo {author}
  {\bibfnamefont {D.}~\bibnamefont {Mandrus}}, \bibinfo {author} {\bibfnamefont
  {M.~D.}\ \bibnamefont {Lumsden}}, \ and\ \bibinfo {author} {\bibfnamefont
  {A.~D.}\ \bibnamefont {Christianson}},\ }\href {\doibase
  10.1103/PhysRevB.92.165128} {\bibfield  {journal} {\bibinfo  {journal} {Phys.
  Rev. B}\ }\textbf {\bibinfo {volume} {92}},\ \bibinfo {pages} {165128}
  (\bibinfo {year} {2015})}\BibitemShut {NoStop}%
\bibitem [{\citenamefont {Brouet}\ \emph {et~al.}(2021)\citenamefont {Brouet},
  \citenamefont {Foulquier}, \citenamefont {Louat}, \citenamefont {Bertran},
  \citenamefont {Le~F\`evre}, \citenamefont {Rault},\ and\ \citenamefont
  {Colson}}]{BrouetPRB21}%
  \BibitemOpen
  \bibfield  {author} {\bibinfo {author} {\bibfnamefont {V.}~\bibnamefont
  {Brouet}}, \bibinfo {author} {\bibfnamefont {P.}~\bibnamefont {Foulquier}},
  \bibinfo {author} {\bibfnamefont {A.}~\bibnamefont {Louat}}, \bibinfo
  {author} {\bibfnamefont {F.~m.~c.}\ \bibnamefont {Bertran}}, \bibinfo
  {author} {\bibfnamefont {P.}~\bibnamefont {Le~F\`evre}}, \bibinfo {author}
  {\bibfnamefont {J.~E.}\ \bibnamefont {Rault}}, \ and\ \bibinfo {author}
  {\bibfnamefont {D.}~\bibnamefont {Colson}},\ }\href {\doibase
  10.1103/PhysRevB.104.L121104} {\bibfield  {journal} {\bibinfo  {journal}
  {Phys. Rev. B}\ }\textbf {\bibinfo {volume} {104}},\ \bibinfo {pages}
  {L121104} (\bibinfo {year} {2021})}\BibitemShut {NoStop}%
\bibitem [{\citenamefont {Moutenet}\ \emph {et~al.}(2018)\citenamefont
  {Moutenet}, \citenamefont {Georges},\ and\ \citenamefont
  {Ferrero}}]{MoutenetPRB18}%
  \BibitemOpen
  \bibfield  {author} {\bibinfo {author} {\bibfnamefont {A.}~\bibnamefont
  {Moutenet}}, \bibinfo {author} {\bibfnamefont {A.}~\bibnamefont {Georges}}, \
  and\ \bibinfo {author} {\bibfnamefont {M.}~\bibnamefont {Ferrero}},\ }\href
  {\doibase 10.1103/PhysRevB.97.155109} {\bibfield  {journal} {\bibinfo
  {journal} {Phys. Rev. B}\ }\textbf {\bibinfo {volume} {97}},\ \bibinfo
  {pages} {155109} (\bibinfo {year} {2018})}\BibitemShut {NoStop}%
\bibitem [{\citenamefont {Zhang}\ \emph {et~al.}(2013)\citenamefont {Zhang},
  \citenamefont {Haule},\ and\ \citenamefont {Vanderbilt}}]{ZhangPRL13}%
  \BibitemOpen
  \bibfield  {author} {\bibinfo {author} {\bibfnamefont {H.}~\bibnamefont
  {Zhang}}, \bibinfo {author} {\bibfnamefont {K.}~\bibnamefont {Haule}}, \ and\
  \bibinfo {author} {\bibfnamefont {D.}~\bibnamefont {Vanderbilt}},\ }\href
  {\doibase 10.1103/PhysRevLett.111.246402} {\bibfield  {journal} {\bibinfo
  {journal} {Phys. Rev. Lett.}\ }\textbf {\bibinfo {volume} {111}},\ \bibinfo
  {pages} {246402} (\bibinfo {year} {2013})}\BibitemShut {NoStop}%
\bibitem [{\citenamefont {Jeong}\ \emph {et~al.}(2020)\citenamefont {Jeong},
  \citenamefont {Lenz}, \citenamefont {Gukasov}, \citenamefont {Fabr\`eges},
  \citenamefont {Sazonov}, \citenamefont {Hutanu}, \citenamefont {Louat},
  \citenamefont {Bounoua}, \citenamefont {Martins}, \citenamefont {Biermann},
  \citenamefont {Brouet}, \citenamefont {Sidis},\ and\ \citenamefont
  {Bourges}}]{JeongPRL20}%
  \BibitemOpen
  \bibfield  {author} {\bibinfo {author} {\bibfnamefont {J.}~\bibnamefont
  {Jeong}}, \bibinfo {author} {\bibfnamefont {B.}~\bibnamefont {Lenz}},
  \bibinfo {author} {\bibfnamefont {A.}~\bibnamefont {Gukasov}}, \bibinfo
  {author} {\bibfnamefont {X.}~\bibnamefont {Fabr\`eges}}, \bibinfo {author}
  {\bibfnamefont {A.}~\bibnamefont {Sazonov}}, \bibinfo {author} {\bibfnamefont
  {V.}~\bibnamefont {Hutanu}}, \bibinfo {author} {\bibfnamefont
  {A.}~\bibnamefont {Louat}}, \bibinfo {author} {\bibfnamefont
  {D.}~\bibnamefont {Bounoua}}, \bibinfo {author} {\bibfnamefont
  {C.}~\bibnamefont {Martins}}, \bibinfo {author} {\bibfnamefont
  {S.}~\bibnamefont {Biermann}}, \bibinfo {author} {\bibfnamefont
  {V.}~\bibnamefont {Brouet}}, \bibinfo {author} {\bibfnamefont
  {Y.}~\bibnamefont {Sidis}}, \ and\ \bibinfo {author} {\bibfnamefont
  {P.}~\bibnamefont {Bourges}},\ }\href {\doibase
  10.1103/PhysRevLett.125.097202} {\bibfield  {journal} {\bibinfo  {journal}
  {Phys. Rev. Lett.}\ }\textbf {\bibinfo {volume} {125}},\ \bibinfo {pages}
  {097202} (\bibinfo {year} {2020})}\BibitemShut {NoStop}%
\bibitem [{\citenamefont {Georges}\ \emph {et~al.}(1996)\citenamefont
  {Georges}, \citenamefont {Kotliar}, \citenamefont {Krauth},\ and\
  \citenamefont {Rozenberg}}]{GeorgesRMP96}%
  \BibitemOpen
  \bibfield  {author} {\bibinfo {author} {\bibfnamefont {A.}~\bibnamefont
  {Georges}}, \bibinfo {author} {\bibfnamefont {G.}~\bibnamefont {Kotliar}},
  \bibinfo {author} {\bibfnamefont {W.}~\bibnamefont {Krauth}}, \ and\ \bibinfo
  {author} {\bibfnamefont {M.~J.}\ \bibnamefont {Rozenberg}},\ }\href@noop {}
  {\bibfield  {journal} {\bibinfo  {journal} {Rev. Mod. Phys.}\ }\textbf
  {\bibinfo {volume} {68}},\ \bibinfo {pages} {13} (\bibinfo {year}
  {1996})}\BibitemShut {NoStop}%
\bibitem [{\citenamefont {Gull}\ \emph {et~al.}(2011)\citenamefont {Gull},
  \citenamefont {Millis}, \citenamefont {Lichtenstein}, \citenamefont
  {Rubtsov}, \citenamefont {Troyer},\ and\ \citenamefont {Werner}}]{GullRMP11}%
  \BibitemOpen
  \bibfield  {author} {\bibinfo {author} {\bibfnamefont {E.}~\bibnamefont
  {Gull}}, \bibinfo {author} {\bibfnamefont {A.~J.}\ \bibnamefont {Millis}},
  \bibinfo {author} {\bibfnamefont {A.~I.}\ \bibnamefont {Lichtenstein}},
  \bibinfo {author} {\bibfnamefont {A.~N.}\ \bibnamefont {Rubtsov}}, \bibinfo
  {author} {\bibfnamefont {M.}~\bibnamefont {Troyer}}, \ and\ \bibinfo {author}
  {\bibfnamefont {P.}~\bibnamefont {Werner}},\ }\href {\doibase
  10.1103/RevModPhys.83.349} {\bibfield  {journal} {\bibinfo  {journal} {Rev.
  Mod. Phys.}\ }\textbf {\bibinfo {volume} {83}},\ \bibinfo {pages} {349}
  (\bibinfo {year} {2011})}\BibitemShut {NoStop}%
\bibitem [{\citenamefont {Levy}\ \emph {et~al.}(2017)\citenamefont {Levy},
  \citenamefont {LeBlanc},\ and\ \citenamefont {Gull}}]{Levy17}%
  \BibitemOpen
  \bibfield  {author} {\bibinfo {author} {\bibfnamefont {R.}~\bibnamefont
  {Levy}}, \bibinfo {author} {\bibfnamefont {J.}~\bibnamefont {LeBlanc}}, \
  and\ \bibinfo {author} {\bibfnamefont {E.}~\bibnamefont {Gull}},\ }\href
  {\doibase https://doi.org/10.1016/j.cpc.2017.01.018} {\bibfield  {journal}
  {\bibinfo  {journal} {Computer Physics Communications}\ }\textbf {\bibinfo
  {volume} {215}},\ \bibinfo {pages} {149} (\bibinfo {year}
  {2017})}\BibitemShut {NoStop}%
\bibitem [{\citenamefont {Hogan}\ \emph {et~al.}(2016)\citenamefont {Hogan},
  \citenamefont {Bjaalie}, \citenamefont {Zhao}, \citenamefont {Belvin},
  \citenamefont {Wang}, \citenamefont {Van~de Walle}, \citenamefont {Hsieh},\
  and\ \citenamefont {Wilson}}]{HoganPRB16}%
  \BibitemOpen
  \bibfield  {author} {\bibinfo {author} {\bibfnamefont {T.}~\bibnamefont
  {Hogan}}, \bibinfo {author} {\bibfnamefont {L.}~\bibnamefont {Bjaalie}},
  \bibinfo {author} {\bibfnamefont {L.}~\bibnamefont {Zhao}}, \bibinfo {author}
  {\bibfnamefont {C.}~\bibnamefont {Belvin}}, \bibinfo {author} {\bibfnamefont
  {X.}~\bibnamefont {Wang}}, \bibinfo {author} {\bibfnamefont {C.~G.}\
  \bibnamefont {Van~de Walle}}, \bibinfo {author} {\bibfnamefont
  {D.}~\bibnamefont {Hsieh}}, \ and\ \bibinfo {author} {\bibfnamefont {S.~D.}\
  \bibnamefont {Wilson}},\ }\href {\doibase 10.1103/PhysRevB.93.134110}
  {\bibfield  {journal} {\bibinfo  {journal} {Phys. Rev. B}\ }\textbf {\bibinfo
  {volume} {93}},\ \bibinfo {pages} {134110} (\bibinfo {year}
  {2016})}\BibitemShut {NoStop}%
\bibitem [{\citenamefont {Louat}\ \emph {et~al.}(2018)\citenamefont {Louat},
  \citenamefont {Bert}, \citenamefont {Serrier-Garcia}, \citenamefont
  {Bertran}, \citenamefont {Le~F\`evre}, \citenamefont {Rault},\ and\
  \citenamefont {Brouet}}]{LouatPRB18}%
  \BibitemOpen
  \bibfield  {author} {\bibinfo {author} {\bibfnamefont {A.}~\bibnamefont
  {Louat}}, \bibinfo {author} {\bibfnamefont {F.}~\bibnamefont {Bert}},
  \bibinfo {author} {\bibfnamefont {L.}~\bibnamefont {Serrier-Garcia}},
  \bibinfo {author} {\bibfnamefont {F.}~\bibnamefont {Bertran}}, \bibinfo
  {author} {\bibfnamefont {P.}~\bibnamefont {Le~F\`evre}}, \bibinfo {author}
  {\bibfnamefont {J.}~\bibnamefont {Rault}}, \ and\ \bibinfo {author}
  {\bibfnamefont {V.}~\bibnamefont {Brouet}},\ }\href {\doibase
  10.1103/PhysRevB.97.161109} {\bibfield  {journal} {\bibinfo  {journal} {Phys.
  Rev. B}\ }\textbf {\bibinfo {volume} {97}},\ \bibinfo {pages} {161109}
  (\bibinfo {year} {2018})}\BibitemShut {NoStop}%
\bibitem [{\citenamefont {Kim}\ \emph {et~al.}(2002)\citenamefont {Kim},
  \citenamefont {Ronning}, \citenamefont {Damascelli}, \citenamefont {Feng},
  \citenamefont {Shen}, \citenamefont {Wells}, \citenamefont {Kim},
  \citenamefont {Birgeneau}, \citenamefont {Kastner}, \citenamefont {Miller},
  \citenamefont {Eisaki},\ and\ \citenamefont {Uchida}}]{KimPRB02}%
  \BibitemOpen
  \bibfield  {author} {\bibinfo {author} {\bibfnamefont {C.}~\bibnamefont
  {Kim}}, \bibinfo {author} {\bibfnamefont {F.}~\bibnamefont {Ronning}},
  \bibinfo {author} {\bibfnamefont {A.}~\bibnamefont {Damascelli}}, \bibinfo
  {author} {\bibfnamefont {D.~L.}\ \bibnamefont {Feng}}, \bibinfo {author}
  {\bibfnamefont {Z.-X.}\ \bibnamefont {Shen}}, \bibinfo {author}
  {\bibfnamefont {B.~O.}\ \bibnamefont {Wells}}, \bibinfo {author}
  {\bibfnamefont {Y.~J.}\ \bibnamefont {Kim}}, \bibinfo {author} {\bibfnamefont
  {R.~J.}\ \bibnamefont {Birgeneau}}, \bibinfo {author} {\bibfnamefont {M.~A.}\
  \bibnamefont {Kastner}}, \bibinfo {author} {\bibfnamefont {L.~L.}\
  \bibnamefont {Miller}}, \bibinfo {author} {\bibfnamefont {H.}~\bibnamefont
  {Eisaki}}, \ and\ \bibinfo {author} {\bibfnamefont {S.}~\bibnamefont
  {Uchida}},\ }\href {\doibase 10.1103/PhysRevB.65.174516} {\bibfield
  {journal} {\bibinfo  {journal} {Phys. Rev. B}\ }\textbf {\bibinfo {volume}
  {65}},\ \bibinfo {pages} {174516} (\bibinfo {year} {2002})}\BibitemShut
  {NoStop}%
\bibitem [{\citenamefont {Dessau}\ \emph {et~al.}(1998)\citenamefont {Dessau},
  \citenamefont {Saitoh}, \citenamefont {Park}, \citenamefont {Shen},
  \citenamefont {Villella}, \citenamefont {Hamada}, \citenamefont {Moritomo},\
  and\ \citenamefont {Tokura}}]{DessauPRL98}%
  \BibitemOpen
  \bibfield  {author} {\bibinfo {author} {\bibfnamefont {D.~S.}\ \bibnamefont
  {Dessau}}, \bibinfo {author} {\bibfnamefont {T.}~\bibnamefont {Saitoh}},
  \bibinfo {author} {\bibfnamefont {C.-H.}\ \bibnamefont {Park}}, \bibinfo
  {author} {\bibfnamefont {Z.-X.}\ \bibnamefont {Shen}}, \bibinfo {author}
  {\bibfnamefont {P.}~\bibnamefont {Villella}}, \bibinfo {author}
  {\bibfnamefont {N.}~\bibnamefont {Hamada}}, \bibinfo {author} {\bibfnamefont
  {Y.}~\bibnamefont {Moritomo}}, \ and\ \bibinfo {author} {\bibfnamefont
  {Y.}~\bibnamefont {Tokura}},\ }\href {\doibase 10.1103/PhysRevLett.81.192}
  {\bibfield  {journal} {\bibinfo  {journal} {Phys. Rev. Lett.}\ }\textbf
  {\bibinfo {volume} {81}},\ \bibinfo {pages} {192} (\bibinfo {year}
  {1998})}\BibitemShut {NoStop}%
\bibitem [{\citenamefont {Perfetti}\ \emph {et~al.}(2001)\citenamefont
  {Perfetti}, \citenamefont {Berger}, \citenamefont {Reginelli}, \citenamefont
  {Degiorgi}, \citenamefont {H\"ochst}, \citenamefont {Voit}, \citenamefont
  {Margaritondo},\ and\ \citenamefont {Grioni}}]{PerfettiPRL01}%
  \BibitemOpen
  \bibfield  {author} {\bibinfo {author} {\bibfnamefont {L.}~\bibnamefont
  {Perfetti}}, \bibinfo {author} {\bibfnamefont {H.}~\bibnamefont {Berger}},
  \bibinfo {author} {\bibfnamefont {A.}~\bibnamefont {Reginelli}}, \bibinfo
  {author} {\bibfnamefont {L.}~\bibnamefont {Degiorgi}}, \bibinfo {author}
  {\bibfnamefont {H.}~\bibnamefont {H\"ochst}}, \bibinfo {author}
  {\bibfnamefont {J.}~\bibnamefont {Voit}}, \bibinfo {author} {\bibfnamefont
  {G.}~\bibnamefont {Margaritondo}}, \ and\ \bibinfo {author} {\bibfnamefont
  {M.}~\bibnamefont {Grioni}},\ }\href {\doibase 10.1103/PhysRevLett.87.216404}
  {\bibfield  {journal} {\bibinfo  {journal} {Phys. Rev. Lett.}\ }\textbf
  {\bibinfo {volume} {87}},\ \bibinfo {pages} {216404} (\bibinfo {year}
  {2001})}\BibitemShut {NoStop}%
\bibitem [{\citenamefont {Shen}\ \emph {et~al.}(2004)\citenamefont {Shen},
  \citenamefont {Ronning}, \citenamefont {Lu}, \citenamefont {Lee},
  \citenamefont {Ingle}, \citenamefont {Meevasana}, \citenamefont {Baumberger},
  \citenamefont {Damascelli}, \citenamefont {Armitage}, \citenamefont {Miller},
  \citenamefont {Kohsaka}, \citenamefont {Azuma}, \citenamefont {Takano},
  \citenamefont {Takagi},\ and\ \citenamefont {Shen}}]{KMShenPRL04}%
  \BibitemOpen
  \bibfield  {author} {\bibinfo {author} {\bibfnamefont {K.~M.}\ \bibnamefont
  {Shen}}, \bibinfo {author} {\bibfnamefont {F.}~\bibnamefont {Ronning}},
  \bibinfo {author} {\bibfnamefont {D.~H.}\ \bibnamefont {Lu}}, \bibinfo
  {author} {\bibfnamefont {W.~S.}\ \bibnamefont {Lee}}, \bibinfo {author}
  {\bibfnamefont {N.~J.~C.}\ \bibnamefont {Ingle}}, \bibinfo {author}
  {\bibfnamefont {W.}~\bibnamefont {Meevasana}}, \bibinfo {author}
  {\bibfnamefont {F.}~\bibnamefont {Baumberger}}, \bibinfo {author}
  {\bibfnamefont {A.}~\bibnamefont {Damascelli}}, \bibinfo {author}
  {\bibfnamefont {N.~P.}\ \bibnamefont {Armitage}}, \bibinfo {author}
  {\bibfnamefont {L.~L.}\ \bibnamefont {Miller}}, \bibinfo {author}
  {\bibfnamefont {Y.}~\bibnamefont {Kohsaka}}, \bibinfo {author} {\bibfnamefont
  {M.}~\bibnamefont {Azuma}}, \bibinfo {author} {\bibfnamefont
  {M.}~\bibnamefont {Takano}}, \bibinfo {author} {\bibfnamefont
  {H.}~\bibnamefont {Takagi}}, \ and\ \bibinfo {author} {\bibfnamefont {Z.-X.}\
  \bibnamefont {Shen}},\ }\href@noop {} {\bibfield  {journal} {\bibinfo
  {journal} {Phys. Rev. Lett.}\ }\textbf {\bibinfo {volume} {93}},\ \bibinfo
  {pages} {267002} (\bibinfo {year} {2004})}\BibitemShut {NoStop}%
\bibitem [{\citenamefont {Gretarsson}\ \emph {et~al.}(2017)\citenamefont
  {Gretarsson}, \citenamefont {Sauceda}, \citenamefont {Sung}, \citenamefont
  {H\"oppner}, \citenamefont {Minola}, \citenamefont {Kim}, \citenamefont
  {Keimer},\ and\ \citenamefont {Le~Tacon}}]{GretarssonPRB17}%
  \BibitemOpen
  \bibfield  {author} {\bibinfo {author} {\bibfnamefont {H.}~\bibnamefont
  {Gretarsson}}, \bibinfo {author} {\bibfnamefont {J.}~\bibnamefont {Sauceda}},
  \bibinfo {author} {\bibfnamefont {N.~H.}\ \bibnamefont {Sung}}, \bibinfo
  {author} {\bibfnamefont {M.}~\bibnamefont {H\"oppner}}, \bibinfo {author}
  {\bibfnamefont {M.}~\bibnamefont {Minola}}, \bibinfo {author} {\bibfnamefont
  {B.~J.}\ \bibnamefont {Kim}}, \bibinfo {author} {\bibfnamefont
  {B.}~\bibnamefont {Keimer}}, \ and\ \bibinfo {author} {\bibfnamefont
  {M.}~\bibnamefont {Le~Tacon}},\ }\href {\doibase 10.1103/PhysRevB.96.115138}
  {\bibfield  {journal} {\bibinfo  {journal} {Phys. Rev. B}\ }\textbf {\bibinfo
  {volume} {96}},\ \bibinfo {pages} {115138} (\bibinfo {year}
  {2017})}\BibitemShut {NoStop}%
\bibitem [{\citenamefont {He}\ \emph {et~al.}(2015)\citenamefont {He},
  \citenamefont {Hogan}, \citenamefont {Mion}, \citenamefont {Hafiz},
  \citenamefont {He}, \citenamefont {Denlinger}, \citenamefont {Mo},
  \citenamefont {Dhital}, \citenamefont {Chen}, \citenamefont {Lin},
  \citenamefont {Zhang}, \citenamefont {Hashimoto}, \citenamefont {Pan},
  \citenamefont {Lu}, \citenamefont {Arita}, \citenamefont {Shimada},
  \citenamefont {Markiewicz}, \citenamefont {Wang}, \citenamefont {Kempa},
  \citenamefont {Naughton}, \citenamefont {Bansil}, \citenamefont {Wilson},\
  and\ \citenamefont {He}}]{HeNatMat15}%
  \BibitemOpen
  \bibfield  {author} {\bibinfo {author} {\bibfnamefont {J.}~\bibnamefont
  {He}}, \bibinfo {author} {\bibfnamefont {T.}~\bibnamefont {Hogan}}, \bibinfo
  {author} {\bibfnamefont {T.~R.}\ \bibnamefont {Mion}}, \bibinfo {author}
  {\bibfnamefont {H.}~\bibnamefont {Hafiz}}, \bibinfo {author} {\bibfnamefont
  {Y.}~\bibnamefont {He}}, \bibinfo {author} {\bibfnamefont {J.~D.}\
  \bibnamefont {Denlinger}}, \bibinfo {author} {\bibfnamefont {S.-K.}\
  \bibnamefont {Mo}}, \bibinfo {author} {\bibfnamefont {C.}~\bibnamefont
  {Dhital}}, \bibinfo {author} {\bibfnamefont {X.}~\bibnamefont {Chen}},
  \bibinfo {author} {\bibfnamefont {Q.}~\bibnamefont {Lin}}, \bibinfo {author}
  {\bibfnamefont {Y.}~\bibnamefont {Zhang}}, \bibinfo {author} {\bibfnamefont
  {M.}~\bibnamefont {Hashimoto}}, \bibinfo {author} {\bibfnamefont
  {H.}~\bibnamefont {Pan}}, \bibinfo {author} {\bibfnamefont {D.~H.}\
  \bibnamefont {Lu}}, \bibinfo {author} {\bibfnamefont {M.}~\bibnamefont
  {Arita}}, \bibinfo {author} {\bibfnamefont {K.}~\bibnamefont {Shimada}},
  \bibinfo {author} {\bibfnamefont {R.~S.}\ \bibnamefont {Markiewicz}},
  \bibinfo {author} {\bibfnamefont {Z.}~\bibnamefont {Wang}}, \bibinfo {author}
  {\bibfnamefont {K.}~\bibnamefont {Kempa}}, \bibinfo {author} {\bibfnamefont
  {M.~J.}\ \bibnamefont {Naughton}}, \bibinfo {author} {\bibfnamefont
  {A.}~\bibnamefont {Bansil}}, \bibinfo {author} {\bibfnamefont {S.~D.}\
  \bibnamefont {Wilson}}, \ and\ \bibinfo {author} {\bibfnamefont {R.-H.}\
  \bibnamefont {He}},\ }\href {http://dx.doi.org/10.1038/nmat4273} {\bibfield
  {journal} {\bibinfo  {journal} {Nature Materials}\ }\textbf {\bibinfo
  {volume} {14}},\ \bibinfo {pages} {577 EP } (\bibinfo {year}
  {2015})}\BibitemShut {NoStop}%
\bibitem [{\citenamefont {Kovaleva}\ \emph {et~al.}(2004)\citenamefont
  {Kovaleva}, \citenamefont {Boris}, \citenamefont {Bernhard}, \citenamefont
  {Kulakov}, \citenamefont {Pimenov}, \citenamefont {Balbashov}, \citenamefont
  {Khaliullin},\ and\ \citenamefont {Keimer}}]{KovalevaPRL04}%
  \BibitemOpen
  \bibfield  {author} {\bibinfo {author} {\bibfnamefont {N.~N.}\ \bibnamefont
  {Kovaleva}}, \bibinfo {author} {\bibfnamefont {A.~V.}\ \bibnamefont {Boris}},
  \bibinfo {author} {\bibfnamefont {C.}~\bibnamefont {Bernhard}}, \bibinfo
  {author} {\bibfnamefont {A.}~\bibnamefont {Kulakov}}, \bibinfo {author}
  {\bibfnamefont {A.}~\bibnamefont {Pimenov}}, \bibinfo {author} {\bibfnamefont
  {A.~M.}\ \bibnamefont {Balbashov}}, \bibinfo {author} {\bibfnamefont
  {G.}~\bibnamefont {Khaliullin}}, \ and\ \bibinfo {author} {\bibfnamefont
  {B.}~\bibnamefont {Keimer}},\ }\href {\doibase 10.1103/PhysRevLett.93.147204}
  {\bibfield  {journal} {\bibinfo  {journal} {Phys. Rev. Lett.}\ }\textbf
  {\bibinfo {volume} {93}},\ \bibinfo {pages} {147204} (\bibinfo {year}
  {2004})}\BibitemShut {NoStop}%
\bibitem [{\citenamefont {Gorelov}\ \emph {et~al.}(2010)\citenamefont
  {Gorelov}, \citenamefont {Karolak}, \citenamefont {Wehling}, \citenamefont
  {Lechermann}, \citenamefont {Lichtenstein},\ and\ \citenamefont
  {Pavarini}}]{GorelovPRL10}%
  \BibitemOpen
  \bibfield  {author} {\bibinfo {author} {\bibfnamefont {E.}~\bibnamefont
  {Gorelov}}, \bibinfo {author} {\bibfnamefont {M.}~\bibnamefont {Karolak}},
  \bibinfo {author} {\bibfnamefont {T.~O.}\ \bibnamefont {Wehling}}, \bibinfo
  {author} {\bibfnamefont {F.}~\bibnamefont {Lechermann}}, \bibinfo {author}
  {\bibfnamefont {A.~I.}\ \bibnamefont {Lichtenstein}}, \ and\ \bibinfo
  {author} {\bibfnamefont {E.}~\bibnamefont {Pavarini}},\ }\href {\doibase
  10.1103/PhysRevLett.104.226401} {\bibfield  {journal} {\bibinfo  {journal}
  {Phys. Rev. Lett.}\ }\textbf {\bibinfo {volume} {104}},\ \bibinfo {pages}
  {226401} (\bibinfo {year} {2010})}\BibitemShut {NoStop}%
\bibitem [{\citenamefont {Wang}\ \emph {et~al.}(2018)\citenamefont {Wang},
  \citenamefont {Okada}, \citenamefont {O{\textquoteright}Neal}, \citenamefont
  {Zhou}, \citenamefont {Walkup}, \citenamefont {Dhital}, \citenamefont
  {Hogan}, \citenamefont {Clancy}, \citenamefont {Kim}, \citenamefont {Hu},
  \citenamefont {Santos}, \citenamefont {Wilson}, \citenamefont {Trivedi},\
  and\ \citenamefont {Madhavan}}]{WangPNAS18}%
  \BibitemOpen
  \bibfield  {author} {\bibinfo {author} {\bibfnamefont {Z.}~\bibnamefont
  {Wang}}, \bibinfo {author} {\bibfnamefont {Y.}~\bibnamefont {Okada}},
  \bibinfo {author} {\bibfnamefont {J.}~\bibnamefont {O{\textquoteright}Neal}},
  \bibinfo {author} {\bibfnamefont {W.}~\bibnamefont {Zhou}}, \bibinfo {author}
  {\bibfnamefont {D.}~\bibnamefont {Walkup}}, \bibinfo {author} {\bibfnamefont
  {C.}~\bibnamefont {Dhital}}, \bibinfo {author} {\bibfnamefont
  {T.}~\bibnamefont {Hogan}}, \bibinfo {author} {\bibfnamefont
  {P.}~\bibnamefont {Clancy}}, \bibinfo {author} {\bibfnamefont {Y.-J.}\
  \bibnamefont {Kim}}, \bibinfo {author} {\bibfnamefont {Y.~F.}\ \bibnamefont
  {Hu}}, \bibinfo {author} {\bibfnamefont {L.~H.}\ \bibnamefont {Santos}},
  \bibinfo {author} {\bibfnamefont {S.~D.}\ \bibnamefont {Wilson}}, \bibinfo
  {author} {\bibfnamefont {N.}~\bibnamefont {Trivedi}}, \ and\ \bibinfo
  {author} {\bibfnamefont {V.}~\bibnamefont {Madhavan}},\ }\href {\doibase
  10.1073/pnas.1808056115} {\bibfield  {journal} {\bibinfo  {journal}
  {Proceedings of the National Academy of Sciences}\ }\textbf {\bibinfo
  {volume} {115}},\ \bibinfo {pages} {11198} (\bibinfo {year} {2018})},\
  \Eprint
  {http://arxiv.org/abs/https://www.pnas.org/content/115/44/11198.full.pdf}
  {https://www.pnas.org/content/115/44/11198.full.pdf} \BibitemShut {NoStop}%
\end{thebibliography}%

\end{document}